\documentclass[aps,pre,showpacs,twocolumn]{revtex4}
\usepackage{epsfig,graphicx}
\usepackage{bm}
\usepackage{epsfig}
\usepackage{amsmath,amssymb,amsfonts,wasysym,color}
\def\be{\begin{equation}}
\def\ee{\end{equation}}
\def\bea{\begin{eqnarray}}
\def\eea{\end{eqnarray}}

\begin{document}

\title{Interplay of complete wetting, critical adsorption, and capillary condensation.}

\author{A. Drzewi\'nski,$^{1}$ A. Macio\l ek,$^{2,3}$$^{,4}$ A. Barasi\'nski,$^{1}$ and S. Dietrich$^{2,3}$}

\affiliation{$^{1}$Institute of Physics, University of Zielona G\'ora, ul. Prof. Z. Szafrana 4a, 
65-516 Zielona G\' ora, Poland}

\affiliation{$^{2}$Max-Planck-Institut f{\"u}r Metallforschung, Heisenbergstr.~3, D-70569 Stuttgart, Germany}
\affiliation{$^{3}$Institut f{\"u}r Theoretische und Angewandte Physik, Universit{\"a}t Stuttgart,
 Pfaffenwaldring 57, D-70569 Stuttgart, Germany}
\affiliation{$^{4}$Institute of Physical Chemistry, Polish Academy of Sciences, Department III, Kasprzaka 44/52,
            PL-01-224 Warsaw, Poland}

\begin{abstract}
The excess adsorption $\Gamma $ in two-dimensional Ising strips $(\infty \times L)$
subject to identical boundary fields, at both one-dimensional surfaces
decaying in the orthogonal direction $j$ as $-h_1j^{-p}$, is studied for various values of $p$
and along various thermodynamic  paths  below the critical point
by means of the density-matrix renormalization-group method.
 The crossover behavior  between the complete wetting and  critical adsorption
regimes, occurring  in  semi-infinite systems, are strongly influenced by 
confinement effects. Along isotherms $T=const$ the asymptotic power law dependences
on the external bulk field, which  characterize these two  regimes, are 
undercut by capillary condensation. 
Along the pseudo first-order phase coexistence line
of the strips, which varies with  temperature, 
 we find a  broad crossover regime where
both  the thickness of the  wetting film and  $\Gamma$
increase as function of the reduced temperature $\tau$  but do not 
follow any  power law.
Above the wetting temperature the order parameter 
 profiles are not slab-like but 
exhibit  wide interfacial variations and pronounced tails.
Inter alia, our explicit calculations demonstrate that, 
contrary to opposite claims by
Kroll and Lipowsky  [Phys. Rev. B {\bf 28}, 5273 (1983)],
  for $p=2$  critical wetting
transitions do  exist and we determine  the corresponding wetting 
 phase diagram in the  $(h_1,T)$ plane.
\date{\today}
\end{abstract}

\maketitle
\section{Introduction}
\label{sec:1}

Far away from phase boundaries, i.e., deep in the one-phase region, condensed matter is perturbed 
by confining walls only within a thin layer proportional to the bulk correlation length  $\xi$, 
i.e., at most approximately 15 \AA.
This changes drastically if the thermodynamic state of the bulk system is moved towards the boundary 
of phase transitions between the bulk phases. If the phase transition is of {\it second} order 
the bulk correlation length diverges and so-called critical adsorption occurs, i.e., 
the perturbation due to the wall located at $y=0$ penetrates deeply into the bulk, resulting in
an algebraic  divergence of the thickness of the interfacial structure
 $\int_0^{\infty}\left(\rho(y)-\rho_b\right)dy/(\rho(0^+)-\rho_b)\equiv \Gamma/(\rho(0^+)-\rho_b)\to \infty$,
where for a one-component fluid as a paradigmatic case $\rho_b$ is the bulk number density for 
a given temperature $T$ and
chemical potential $\mu$,  and $\rho(y) $ is the fluid number density
profile~\cite{binder,diehl}.
Moreover, the order-parameter profile $m(y)\equiv (\rho(y)-\rho_c)/\rho_c$,
where $\rho_c$ is the bulk critical number density,
is  governed by a  universal scaling function
which decays exponentially  on the length scale of $\xi$ in the direction
$y$ normal to the surface~\cite{binder,diehl}.
If the phase transition is of  {\it first} order 
 wetting phenomena occur as a result of a subtle interplay between the
substrate potential, interactions among fluid particles, and entropic 
contributions, in particular interfacial fluctuations \cite{dietrich}.
The thickness of the interfacial structure diverges too, but here with 
exponents which are determined by the decay exponents of the pair potentials
between the fluid particles and of the substrate potential \cite{dietrich}.
These two types of phenomena
are  of significant practical importance, in particular for the condensed
phase being a fluidum near the gas-liquid transition, such as water near the 
thermodynamic coexistence with its vapor,
or binary liquid mixtures such as water and hydrocarbons near liquid-liquid phase transitions. In 
this context various applications  arise ranging
from the use of  colloidal
suspensions \cite{xia}  to   petroleum recovery \cite{bertrand}. 
Wetting films are relevant in many types of
liquid coating processes, such as  lubrication and adhesion and also
for   microfluidics and nanoprinting \cite{deGennes,tabeling}.
Critical adsorption plays an important role, e.g., for heterogeneous nucleation
in supercritical solvents \cite{supercritical} or in  micro- and 
nanofluidic systems  in order to achieve wetting of these small
structures ~\cite{austin}.

\begin{figure}[hbt]
\centering
\includegraphics[width=7.0cm]{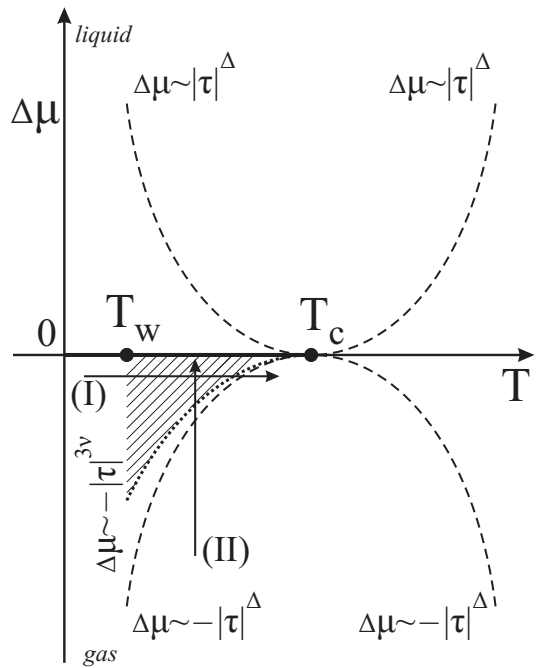}
\caption{Schematic drawing of the liquid-gas coexistence curve $\Delta\mu=0$
in the $(\Delta\mu,T)$ bulk phase diagram.
 The thick solid line indicates the
bulk coexistence line.  $T_w$ is the transition temperature for
a critical wetting transition. Thermodynamic paths (I) and (II), along which
  the behavior of the adsorption is discussed in the main text,
are displayed as well as
various crossover lines.
The hatched area is the complete wetting regime. Note that $3\nu\simeq 1.89$
is larger than $\Delta\simeq 1.56$.}
\label{fig:0}
\end{figure}

A characteristic property of liquids is that  first-order gas-liquid 
or liquid-liquid phase transitions end at critical points.  Therefore 
critical adsorption and wetting must emerge from each other  upon moving along
the first-order phase boundary.
This creates a puzzle. Liquids governed by dispersion forces
(decaying asymptotically $\sim r^{-6}$ with distance  $r$)
belong to the Ising   universality class
\cite{pfeuty}, i.e., the critical exponents describing the singular
behavior of  various thermodynamic  quantities and structural properties 
are those of the Ising universality class. This also holds at interfaces. 
Specifically, along  bulk coexistence  $\Delta \mu=(\mu-\mu_0(T))/(k_BT_c)=0$
one has asymptotically ~\cite{binder,diehl}
\begin{equation}
\label{eq:I_1}
\Gamma \sim |\tau|^{\beta-\nu}, \qquad \Delta \mu=0, 
\end{equation}
 where   $\tau\equiv (T-T_c)/T_c$
is the reduced deviation from the critical temperature $T_c$
and $\Delta \mu $ measures, in units of $k_BT_c$, the deviation
 of the chemical potential
from its value $\mu_0(T)$ at two-phase coexistence, whereas
along the path $\tau=0$
\begin{equation}
\label{eq:I_2}
\Gamma \sim |\Delta \mu|^{(\beta-\nu)/\Delta}, \qquad \tau=0,
\end{equation}
where   the critical exponents  $\nu, \beta$, and $\Delta$ take the
values of the Ising model
(in $d=3$: $\nu=0.6301(4), \beta=0.32653(10)$,
and  $\Delta=3\nu-\beta = 1.564 $ \cite{pelisseto}).
The latter universal  singularity $\Gamma \sim |\Delta \mu|^{-0.194}$ 
is  weaker than the non-universal one  $\Gamma \sim |\Delta \mu|^{-1/3}$ for complete 
wetting \cite{dietrich}, although according to the general
renormalization group (RG) arguments the dispersion forces, 
which are responsible for this non-universal behavior, at first glance
should give rise only to corrections 
to scaling (i.e., subdominant power laws). Naturally the question arises how  the system manages
to restore universality (i.e., dominance by the weaker universal power law)
upon moving  from the non-universal complete wetting 
behavior to the universal critical adsorption behavior.
This issue has been addressed in Refs.~\cite{tarazona:87,dietrich:88:0,TelodaGama:91} and the proposed 
scenario is such that upon approaching bulk coexistence, i.e., for $\tau$ fixed 
and $\Delta \mu\to 0^-$ (see path (II) in  Fig.~\ref{fig:0}), 
the system will {\it always} cross over ultimately to the complete wetting
regime but such that upon approaching $T_c$  (i.e., for smaller $|\tau|$)
this crossover  occurs  closer and closer to the coexistence curve.
The proposed mechanism  is based on the argument that long-ranged dispersion forces 
are relevant and can dominate if  the thickness $\ell \sim |\Delta \mu|^{-1/3}$
 of the wetting layer
is much larger than the bulk correlation length $\xi\sim |\tau|^{-\nu}$. 
As described more closely   in Sec.~\ref{sec:2},  
for dispersion forces in $d=3$, this occurs  
if  $|\Delta \mu| \ll |\tau|^{3\nu}$ 
(see the hatched area in  Fig.~\ref{fig:0}). Accordingly, upon
approaching $T_c$ this complete wetting regime becomes vanishingly small.
This expectation follows similar arguments as  how dispersion forces
become irrelevant for, e.g., two-point correlation functions
in the bulk or for the critical adsorption profile itself, for which RG
theory predicts  exponential decays which, however, 
are dominated asymptotically by the algebraic decay due to the dispersion forces.
Nonetheless the universal scaling functions,
 which capture the exponential decay, are expected to describe also the neighborhood
of $T_c$ correctly, even in the presence of the irrelevant dispersion forces.

So far this puzzle has not been resolved satisfactorily by explicit calculations.  The main reason
is  that mean field theory (MFT) as the natural theoretical starting point 
does not capture 
the correct $d=3$ exponents,
which is essential for disentangling  the different contributions and 
to describe the crossover. Within  MFT, $\nu=\beta$ and
the adsorption diverges  only   $\sim \ln |\Delta \mu|$.
In systems governed by short-ranged forces only, within MFT both critical adsorption and 
complete wetting follow a divergence $\sim \ln|\Delta \mu|$
and therefore they cannot be distinguished; but the divergences differ beyond MFT.
However, going beyond  MFT  is a difficult  theoretical challenge, due to the spatial inhomogeneity.

There is also an urgent experimental need to resolve this issue.
An ellipsometry study of gravity-thinned complete wetting layers 
in the  binary-liquid mixture of cyclohexane and methanol \cite{fenistein:02}
reports results which are  {\it not} in accordance with the theoretical predictions. The data indicate
the divergence of the film thickness
upon  approaching the bulk critical temperature characterized by
a critical exponent which is distinct from the expected, critical exponent
for the bulk correlation length. Scaling arguments
have been put forward  according to which  the observed    {\it effective}
 exponent is associated with  a broad, 
intermediate scaling  regime  facilitating  the crossover
 from complete wetting in the presence of dispersion forces to critical scaling \cite{fenistein:02}.
Also recent  neutron reflectometry data for the adsorption from 
 alkane-perfluoroalkane mixtures at fluorophobic and fluorophilic 
surfaces \cite{bowers:04} are {\it not} in  agreement with the theoretical predictions.
These authors have  found that the behavior of the adsorption, 
 as a function of temperature
 in the one-phase region upon
approaching  liquid-liquid coexistence with the composition 
well removed from the critical composition,  can be represented
by a power law 
with an exponent which differs from both that for complete wetting
and  that for  critical adsorption \cite{bowers:04}. 

 The   predictions described above for the crossover 
between the complete wetting regime of fluids
with long-ranged interactions and the critical regime   are based
on  general scaling arguments and on the
analysis of simple models for the  effective interface Hamiltonian  which are supposed to
describe the relevant physics at length scales much larger than the 
bulk correlation length  \cite{dietrich:88:0}. 
Only  few microscopic  studies are available which test these ideas 
\cite{tarazona:87,TelodaGama:91}.
Those are based on density functional methods and  are of mean-field
character, which in $d=3$ provides a correct description of complete 
wetting in the presence of dispersion forces but
fails in the regime dominated by critical fluctuations.
This makes it even more important to thoroughly 
  analyze  a  model system which takes into account  both the  bulk 
and surface critical fluctuations of semi-infinite systems near $T_c$ 
as well as the interfacial fluctuations associated with wetting phenomena.

As a representative of the corresponding universality class, an
Ising ferromagnet is well suited for such an analysis. In addition, addressing the issue in two spatial 
dimensions ($d=2$) allows one to gain deep  insight because the "exact" solvability provides, 
inter alia, the correct  exponents. 
There is a recent experimental interest in such genuine two-dimensional systems
in the context of proteins immersed in  a fluid  two-component lipid membrane
which is near phase separation, including a critical point belonging to the
$d=2$ Ising universality class \cite{silvius,honerkamp}. If two such proteins are close to each other
 the structural properties of the membrane between
them can be described in terms of the strips studied here, giving even rise to
interesting effective interactions between the proteins.

In the present analysis we use the 
density-matrix renormalization group (DMRG) method
\cite{white:92,nishino,peschel,schollwoeck} to numerically investigate
 the  behavior of the adsorption in a two-dimensional ($d=2$) 
Ising model with short-ranged as well as long-ranged surface fields.
 The  DMRG method provides essentially
exact numerical results for  thermodynamic quantities
and correlation functions, including the magnetization profiles.
 This  allows us to study systematically  the adsorption
 properties along various thermodynamic
paths and thus to test the aforementioned predictions of the mesoscopic, 
effective interface Hamiltonian approaches.
The DMRG method is  based on the transfer matrix approach and it is 
a numerically very efficient  iterative truncation
  algorithm for constructing the effective
transfer matrices for large systems. The method was originally
developed  by White \cite{white:92} for the diagonalization of  quantum spin
chains. It was then adapted by Nishino \cite{nishino} to 
two-dimensional classical systems. 
The DMRG method allows one to study
strips that are infinitely long with widths up to $L=700$ lattice constants 
with  arbitrary surface and bulk fields.
The comparison with  exact results in the case of a  vanishing 
bulk field and in the presence of  contact surface fields shows that the
DMRG method provides  a very high
accuracy for a  broad range of temperatures.

Because the method requires finite values of $L$ the influence
of the distant  surface on the structural properties near the surface under study  cannot be neglected, in particular,
near the critical point --- even for strips  as wide as $L=700$.
On one side, this for the DMRG method unavoidable
 confinement of the  system
complicates the aforementioned crossover behavior 
due to finite-size effects and  capillary condensation. On the other side,
the interplay
between  complete wetting,  critical adsorption, and capillary condensation
do not appear to have been discussed before, although  
it might be of relevance  for
 adsorption-induced colloidal aggregation in binary liquid mixtures 
\cite{beysens,guo}. Our approach is naturally suited for investigating
 interesting and relevant aspects of these important phenomena. 

The  paper is organized as follows. In Sec.~\ref{sec:2}  
theoretical predictions for  semi-infinite systems are summarized.
In Sec.~\ref{sec:3} we describe the microscopic  model and the method. 
Numerical results for the phase diagram of the present
model are described in Sec.~\ref{sec:pd}.
In Sec.~\ref{sec:wtemp} we determine the  wetting temperature
for different ranges of the boundary fields.
Results for  the magnetization  profiles, 
the thickness of the wetting layer, and   the adsorption
along the isotherms and
 along the line of the pseudo-phase coexistence
are reported in Sec.~\ref{sec:res}.
Section~\ref{sec:5} summarizes and concludes our results.

\section{Description of semi-infinite systems}
\label{sec:2}

A useful global characterization of the interfacial structure near a single wall is provided 
by the coverage  $\Gamma$   defined as the  
excess number   of   fluid particles per area adsorbed on the confining  substrate:
\begin{equation}
\label{eq:1}
\Gamma=\int_0^{\infty}(\rho(y)-\rho_b)dy
\end{equation}
where $\rho _b$ is the bulk number density for a  given temperature $T$ 
and  chemical potential $\mu$; here  the fluid number density profile 
$\rho({\bf r})\equiv \rho(y)$ is assumed to vary only in the direction normal
to the wall  located at $y=0$.

\subsection{Complete wetting}
\label{subsec:cw}

If the substrate potential is sufficiently strong there is a wetting transition
temperature $T_w$ such that, if the bulk gas phase approaches 
gas-liquid coexistence $\mu_0(T)$ along isotherms at temperatures $T>T_w$,
$\Gamma$ diverges due to complete wetting, i.e., a
 macroscopically thick wetting film is formed. The equilibrium 
 thickness $\ell_0$ of the 
wetting film can be defined as 
\begin{equation}
\label{eq:el}
\ell_0=\Gamma /(\rho_l-\rho_g),
\end{equation} 
where $\rho_l$ and $\rho_g$
are the bulk number  densities of the liquid and gas phase, respectively, 
at coexistence. In magnetic language the fluid is an Ising ferromagnet,
 the gas phase corresponds to the spin down phase, the liquid phase to the spin up phase, and the difference between the substrate potential and its
analogue for fluid-fluid interactions to a surface field;
the  undersaturation $\Delta \mu=(\mu-\mu_0(T))/(k_BT_c)$
is proportional  to the bulk field $H$.

In the {\it c}omplete 
wetting regime the increase of the adsorption  upon approaching 
  bulk coexistence can be described as \cite{dietrich}  
\begin{equation}
\label{eq:2}
\Gamma (\Delta \mu \to 0,T)\sim |\Delta \mu|^{-\beta_s^{co}}, \quad T_w<T<T_c.
\end{equation}
The exponent $\beta_s^{co}$ for this {\it s}urface quantity 
depends on the form of the fluid-fluid 
and substrate-fluid forces as well as on the spatial dimension $d$.
In  $d=3$,  $\beta_s^{co}$ is non-universal; $\beta_s^{co}=0$ (i.e., $\sim \ln\Delta \mu$) 
for short-ranged forces whilst $\beta_s^{co}=1/p$ for 
wall-fluid and fluid-fluid pair
potentials  decaying  as $r^{-(d+p)}$ ($p=3$ for non-retarded dispersion forces).
Because the upper critical dimension for complete wetting  with
long-ranged forces is $d^{*}_s=3-4/(p+1)<3$ \cite{dietrich,schick,lipowsky} 
these mean-field exponents $\beta_s^{co}=1/p$ 
are not altered  by interfacial fluctuations. 
For short-ranged forces, i.e., $p\to\infty$ one has  $d^{*}_s=3$ 
so that  fluctuations matter in $d=3$, but it turns that for complete wetting they change
 only the amplitude of the thickness of wetting film \cite{dietrich}.

In $d=2$  interfacial fluctuations in the wetting films are
 much stronger and  $2=d<d^{*}_s$ for both short-ranged and dispersion forces
($p=4$ for the latter in $d=2$). Accordingly the complete wetting exponent 
takes a  universal value which turns out to be given by $\beta_s^{co} =1/3$,  
 provided $p\geq 3$ so that fluctuations
dominate \cite{schick,lipowsky}. 
From the point of view of an effective interface Hamiltonian
this latter universality of $\beta_s^{co}$ is due to
the entropic  effects of the unbinding interface  which give rise
 to an  effective repulsive interaction for the gas-liquid interface,
taken to be located on average at $y=\ell$,
which decays $\sim {\ell}^{-\kappa}$ with
 $\kappa=2(d-1)/(3-d)$; $\kappa(d=2)=2$.
If this entropic repulsion  dominates the effective interaction contribution
$\sim {\ell}^{-(p-1)}$, i.e., if $p > \kappa +1$ one finds
$\beta_s^{co}=1/3$ in $d=2$; this defines 
the so-called weak fluctuation regime for complete wetting. 
According to this argument, for $p<3$, one has $\beta_s^{co}=1/p$.
The considerations leading to the above predictions are valid only if the
equilibrium wetting film thickness $\ell_0 $ is much larger than the 
 bulk correlation length $\xi $, i.e., $\ell_0 \gg \xi$.

\subsection{Critical adsorption}
\label{subsec:ca}

Near a critical point $T_c$,  a confining wall generically 
provides an effective
surface field $h_1$ acting on the order parameter (OP)  $m$ describing
 the continuous phase transition and leading to the so-called 
critical adsorption \cite{binder,diehl,fdg}.
The ensuing  decay of the OP profile 
$m(y)\equiv (\rho(y)-\rho_c)/\rho_c$, where $\rho_c$ is the critical
density, follows the power law
\begin{equation}
\label{eq:3}
m(y)\sim y^{-\beta/\nu}, \qquad T=T_c,
\end{equation}
where $\beta$ and $\nu$ are the known  critical exponents of the bulk OP 
and of the correlation length $\xi$, respectively. 
Off the critical point
a crossover to the exponential decay 
\begin{equation}
\label{eq:4}
m(y)\sim \exp(-y/\xi)
\end{equation}
takes place at the distance $y\sim \xi$ from the surface.
This is described by universal scaling functions $P_{\pm}$ such that
$m_{\pm}(y,\tau,\Delta \mu)=m_b^{(0)} P_{\pm}(y/\xi_{\pm}(\tau,\Delta\mu),\Delta \mu|\tau|^{-\Delta})$, 
where $m_b^{(0)}=m(y\to\infty,T<T_c,\Delta\mu=0^{-})=m_0|\tau|^{\beta}<0$, 
$\tau=(T-T_c)/T_c$, and $\Delta$ is the so-called  gap exponent; 
$\pm$ corresponds to $\tau \gtrless 0$.
Accordingly  the adsorption  $\Gamma_{\pm}$ 
diverges as 
\begin{equation}
\label{eq:5}
\Gamma_{\pm} = m_0|\tau|^{\beta}\xi_{\pm}(\tau,\Delta\mu)K_{\pm}(\Delta\mu|\tau|^{-\Delta}),
\end{equation}
where $K_{\pm}(x)=\int_0^{\infty}[P_{\pm}({\tilde y}_{\pm},x)-P_{\pm}(\infty,x)]d{\tilde y}_{\pm}$ \cite{floeter}.
Near the critical point the bulk correlation length acquires the scaling form
\begin{equation}
\label{eq:6}
\xi_{\pm}(\tau, \Delta\mu) = |\tau|^{-\nu}\Xi_{\pm}(\Delta\mu |\tau|^{-\Delta}),
\end{equation}
where  $\Xi_{\pm}$ are  scaling functions. Since 
$\Xi_{\pm}(0)= const =\xi_0^{\pm}$ one has
\be
\label{eq:xitau}
\xi_{\pm}(\tau,0)= \xi_0^{\pm}|\tau|^{-\nu},
\ee
whereas $\Xi_{\pm}(x\to \pm \infty)\sim |x|^{-\nu/\Delta}$ and thus at
 $\tau=0$  
\be
\label{eq:ximu}
\xi(0,\Delta\mu)=\xi_0^{(\mu)}|\Delta\mu|^{-\nu/\Delta}.
\ee
Together with $K_{\pm}(x\to 0)=const$ and $K_{\pm}(x\to \pm\infty)\sim |x|^{\beta/\Delta}$, this 
implies that upon approaching  $(T_c, \mu_0(T_c))$ along the path $\Delta\mu=0$ the adsorption  diverges according to 
 Eq.~(\ref{eq:I_1}) whereas it diverges according to  Eq.~(\ref{eq:I_2}) along the path $\tau=0$.
We note that both Eq.~(\ref{eq:I_1}) and Eq.~(\ref{eq:I_2}) are consistent with the
scaling behavior of $\Gamma\sim m\xi$ due to $m\sim |\tau|^{\beta}$ and
$\xi\sim|\tau|^{-\nu}$ for $\Delta \mu=0$ and due to $m\sim |\Delta\mu|^{1/\delta}$, $\delta=\Delta/\beta$,
 and $\xi\sim |\Delta\mu|^{-\nu/\Delta}$ for $\tau=0$.
Within  MFT  $\beta=\nu=1/2$ which results in  a  logarithmic divergence along both paths.

Equation ~(\ref{eq:I_2}) is expected to hold \cite{dietrich:88:0} also below $T_c$ 
for $|\Delta\mu|\gg |\tau|^{\Delta}$ on both the  wetting  and non-wetting side
 of the coexistence curve, i.e.,  if  the argument 
of the scaling functions  $\Xi_{\pm}$ approaches  infinity  so that   the behavior of
$\xi_{\pm}$ is governed by  $\Delta\mu$ (Eq.~(\ref{eq:ximu})). 
 This regime is referred to as the {\it critical adsorption regime}.
Similarly, above $T_c$ Eq.~(\ref{eq:I_1}) is expected to  be valid  
 within the regime $|\Delta\mu|\ll |\tau|^{\Delta}$, i.e., 
where  the behavior of
$\xi$ is governed by  $\tau$ (Eq.~(\ref{eq:xitau})). 
 The same should hold close to bulk 
coexistence below $T_c$.
However,  it has been  argued 
\cite{dietrich:88:0,tarazona:87,fenistein:02} that even 
near the critical point  long-ranged forces are important,  
in the sense that for $\tau$ fixed and $\Delta \mu\to 0^-$ 
the system will {\it always} cross over to the complete wetting 
regime (Eq.~(\ref{eq:2}))
although upon approaching $T_c$ (i.e., for smaller $\tau$)
 this crossover will occur closer and closer to the
coexistence curve. The expectation for this scenario
to hold  is based on the argument that
long-ranged forces  are relevant and thus dominant
for $\ell_0 \gg \xi$.
Since $\ell_0 \sim |\Delta \mu|^{-\beta_s^{co}}$ and $\xi \sim |\tau|^{-\nu}$,
 this leads to the condition  $|\Delta \mu| \ll |\tau|^{\nu/\beta_s^{co}}$, i.e.,
  $|\Delta \mu| \ll \tau^{3\nu}$ for  dispersion forces in $d=3$. 
Thus on the gas side  region between the coexistence curve $\Delta \mu=0$
and the curve $\Delta \mu =const\times |\tau|^{\nu/\beta_s^{co}}$ 
the adsorption 
 should be governed
by wetting phenomena whereas  the  divergence
of the adsorption $|\tau|^{\beta-\nu}$  according to  Eq.~(\ref{eq:I_1}) 
should be limited to the range
 $|\Delta\mu|^{1/\Delta} \ll |\tau| \ll |\Delta \mu|^{\beta_s^{co}/\nu}$. 
 Furthermore, the  effective interface Hamiltonian approach predicts that
  contrary to the critical adsorption  regime,  in the
wetting dominated region the divergence should  depend on the 
choice of the thermodynamic path taken.  In particular, 
 for any isotherm within this regime
one expects $\Gamma\sim |\Delta\mu|^{-\beta_s^{co}}$,
 but along a  path $\Delta\mu=const|\tau|^x$ with 
$x>\nu/\beta_s^{co}$ one expects  $\Gamma\sim |\tau|^{\beta-x\beta_s^{co}}$.
This follows from  Eq.~(\ref{eq:el}) with $\Delta \rho \sim |\tau|^{\beta}$
 and $\ell_0\sim |\Delta \mu|^{-\beta_s^{co}}$.

\subsection{Crossover phenomena}
\label{subsec:cross}

These regimes described above are expected to  give rise
 to rich crossover phenomena for  the adsorption  $\Gamma (\Delta\mu,T)$ 
upon  crossing boundary lines between them along various thermodynamic paths.
Two of them  are particularly relevant for the present work:
 (i)  isotherms
$T=const<T_c$ with $\Delta \mu\to 0^-$ (see path (II) in Fig.~\ref{fig:0})
 and (ii) a path parallel to the
coexistence curve on the gas side with a small 
undersaturation  $\Delta \mu=const < 0$  and with $T\to T_c$ (path (I) 
in  Fig.~\ref{fig:0}).

Along  an isotherm  $T=const >T_w$ and below
the curve $\Delta \mu \sim -|\tau|^{\Delta}$ (see Fig.~\ref{fig:0})
$\Gamma$ increases  as $|\Delta \mu|^{(\beta-\nu)/\Delta}$ upon
approaching bulk  coexistence from the gas side, i.e., for decreasing
$|\Delta \mu|$, until one enters the crossover
region between the lines  $\Delta\mu=-const |\tau|^{\Delta}$
and $\Delta\mu=-const |\tau|^{\nu/\beta_s^{co}}$ in which $\Gamma$ 
increases further; but therein  no specific and well defined power law can  be expected.
Finally, when the path crosses
the curve $\Delta\mu=-const|\tau|^{\nu/\beta_s^{co}}$ it enters the
 regime governed by
wetting phenomena so that the adsorption should diverge as
 $|\Delta\mu|^{-\beta_s^{co}}$. 
For $\mu=\mu_0$ the wall is wet and the adsorption is
infinite (provided $T_w<T\le T_c$).

The behavior of the adsorption  along a path parallel to the
coexistence curve  is expected 
to be equally rich. Let us consider the case that the wetting transition
at coexistence is continuous at $T=T_w$. According to the theory 
of wetting phenomena \cite{dietrich} the 
adsorption  $\Gamma$ increases smoothly to some finite value (because
$|\Delta\mu| >0$)
 upon approaching the wetting temperature $T_w$. This increase is governed
by    the scaling laws with respect
to $\Delta\mu$ and $\tau$  which are associated with 
critical wetting. Upon a further increase of the
temperature the adsorption  should slightly decrease  because 
$\Gamma=\ell_0 \Delta \rho$, $\ell_0=const$ due to $\Delta\mu =const$, and
$\Delta \rho=\rho_l-\rho_g\sim |\tau|^{\beta}$ until the
crossover line $\Delta\mu=-const|\tau|^{\nu/\beta_s^{co}} $
 to the critical adsorption
regime is reached. There $\Gamma$  should increase  
 $\sim |\tau|^{\beta-\nu}$ until the next  crossover line
$\Delta\mu=-const |\tau|^{\Delta}$ is encountered
beyond which $\Gamma$ as a function of $\tau$ should saturate
at a certain large value $\sim |\Delta\mu|^{(\beta-\nu)/\Delta}$.
Finally, after passing the right branch of the crossover line $\Delta\mu \sim -\tau^{\Delta}$ 
(above $T_c$) the adsorption should decrease $\sim \tau^{\beta-\nu}$ upon increasing $\tau$. 

\begin{figure}[hbt]
\centering
\includegraphics[width=7.5cm]{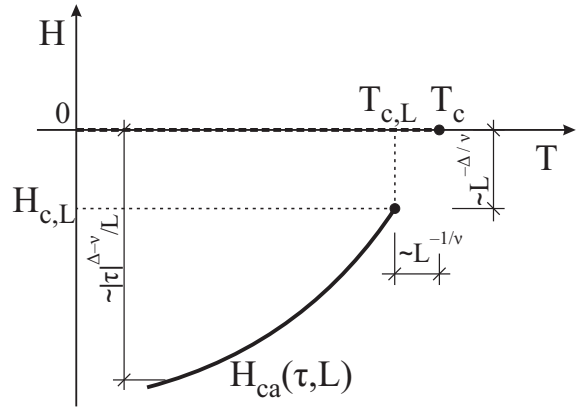}
\caption{ Scaling behavior of the capillary condensation line $H_{ca}(\tau,L)$
near bulk criticality $(T=T_c,H=0)$. For $\tau=(T-T_c)/T_c\ne 0$ and
$L\to \infty$, $H_{ca}(\tau,L)$ approaches bulk coexistence $H=0$
as $|\tau|^{\Delta-\nu}/L$ and the capillary critical point
$(T_{c,L},H_{c,L})$
approaches  $T_c$ and $H=0$ as $L^{-1/\nu}$ and $L^{-\Delta/\nu}$,
respectively. The latter is valid for strong surface field and
takes the form $L^{-(\Delta-\Delta_1)/\nu}$, where $\Delta_1$ is the surface
gap exponent, for weak $h_1$ ~\cite{BLM}. Here $T_{c,L}<T_c$. Note the upward bent of the capillary
condensation line; this feature is due to $\Delta-\nu<1$ (see, c.f., the text
following Eq.~(\ref{eq:capsc})). However, this bent is weak: $\Delta-\nu=7/8=0.875$ in $d=2$ 
and $\Delta-\nu\simeq 0.936$ in $d=3$.
}
\label{fig:12}
\end{figure}

\section{Microscopic model}
\label{sec:3}

In this section we introduce  the microscopic model 
within which we  investigate quantitatively the crossover
regimes described above. 
Specifically, we consider an  Ising ferromagnet
 in a slit geometry subject to identical boundary fields.
Contingent on the type of numerical approach we shall use,  our results
refer to  $d=2$ strips defined on a  square lattice 
of size $M\times L, M\to \infty$. The lattice consists of $L$
 parallel rows at spacing $a$, so that the width of the strip is $La$;
 in the following we set $a=1$. 
Successive rows are labeled by an  index $j$.
At each site, labeled $(k,j)$, there is an Ising 
spin variable taking the value $\sigma_{k,j}=\pm 1$. The boundary surfaces
 are located
in the rows  $j=1$ and $j=L$ and periodic boundary conditions 
(PBCs) are assumed in the lateral  $x$ direction.
Our model Hamiltonian for the  strip with PBCs and $M\to \infty$ is given by

\begin{eqnarray}
\label{eq:3_1}
{\cal H} & = & -J\left( \sum_{\langle kj,k'j'\rangle}\sigma_{j,k}\sigma_{j',k'}\right. \nonumber \\
\mbox{}&\mbox{}&+\left.\sum_{j=1}^{L}V^{ext}_{j,L}\sum_k\sigma_{k,j}+H\sum_{k,j}\sigma_{k,j}\right),
\end{eqnarray}
where the first sum is over all nearest-neighbor pairs and
the external potential is measured in units of $J>0$.
$V^{ext}_{j,L}=V^{s}_{j}+V^{s}_{L+1-j}$ is the total boundary field experienced by a spin in 
the row $j$; it  is the sum of the two independent wall contributions.
The {\it s}ingle-boundary field $V^s_{j}$ is assumed to have the  form
\begin{equation}
\label{eq:3_2}
V^s_{j}=\frac{h_1}{j^p}
\end{equation}
with $p>0$ and $h_1>0$. $H$ is a bulk magnetic field. 
According to Eq.~(\ref{eq:3_1}) $h_1$ and $H$ are dimensionless. 

This model can be viewed as being obtained from 
 a $2d$ lattice gas model mimicking a two-dimensional one-component  fluid
with a short-ranged interaction potential 
between the fluid particles and either short-ranged or long-ranged substrate potentials.
The equivalence between the lattice gas and the Ising model implies the following 
relationships (see, e.g., Ref.~\cite{pandit}): the bulk magnetic field $H$ in the former is proportional 
to the deviation of the chemical potential from the bulk phase boundary
$\mu_0(T)$ in the latter, i.e., $H\sim \Delta \mu$. The lattice-gas 
analogue of the number density in the fluid is related to the magnetization by 
$\rho=(m+1)/2$, so that $\Delta \rho=\rho_l-\rho_g\sim 2m_b$, where
$m_b$ is the spontaneous magnetization. Finally $4J$ corresponds 
to the strength of the attractive pair potential between the fluid
particles, taken to be  short-ranged so that in the lattice gas model 
it can be modelled by nearest neighbor interactions. 
 $V^{ext}$ is a combination of the substrate
potential and the liquid-liquid interaction. 
These relationships can be extended to binary liquid mixtures ~\cite{getta}.

We recall that although there is no longer any true phase transition for finite 
$L$, in two-dimensional Ising strips with large $L$ there is still a line of extremely weakly
 rounded first-order transitions ending at a pseudocritical point the location
of which  in the plane $(H,T)$ spanned by the bulk field
and the   temperature $T$ depends on the character of the surface fields
\cite{fisherprivman,binder:08}. For surfaces which  prefer  the same
bulk  phase, the phenomenon equivalent to capillary condensation takes place.
The pseudo-phase coexistence between phases of spin up and spin down   occurs along a line
$H_{ca}(T,L)$, 
which is given approximately by the analogue of the Kelvin equation
 \cite{kelvin}:
\begin{equation}
\label{eq:Kelvin}
H_{ca}(T,L)\approx -\sigma(T)/\left(L|m_b^{(0)}(T)|\right),
\end{equation}
where $\sigma(T)$  is the interfacial tension (divided by $J$) between the coexisting 
bulk phases and $m_b^{(0)}(T)<0$ is the spontaneous bulk magnetization for $H=0^{-}$.
For the $d=2$ Ising model, the surface tension is given exactly
$\beta J \sigma(T)=2(K-K^*)$, where $K=J/k_BT$ and $K^*$ satisfies
$\sinh 2K\sinh 2K^*=1$.
The occurrence  of  thick wetting films of $+$ spins at the two surfaces,
for a thermodynamic bulk state corresponding to 
$-$ spins, gives rise to nontrivial corrections to Eq.~(\ref{eq:Kelvin})
which  shift the 
condensation line to larger values of $|H|$ \cite{binder:08,evans:90}.
The pseudo-coexistence line ends at  a pseudo-capillary critical
 point $(H_{c,L},T_{c,L})$
where $T_{c,L}(h_1,p)$ lies below the temperature $T_c$ 
of the bulk critical point.
Its position as well as that  of $H_{c,L}$ depends on $L$ and
the strength  and the range of the surface fields.
For large $L$ and strong $h_1$ the  shifts of the critical temperature
and of  the bulk field $H_{c,L}(h_1,p)$ are given by \cite{binder:08,fisher} 
(see, c.f., Fig.~\ref{fig:12})
\begin{equation}
\label{eq:shifts}
T_{c,L}-T_c\sim -L^{-1/\nu}, \quad H_{c,L}\sim -L^{-\Delta/\nu}
\end{equation}
with $\nu=1$ and $\Delta=15/8$ for the two-dimensional Ising model.

\section{Phase diagram}
\label{sec:pd}

\begin{figure}[hbt]
\centering
\includegraphics[width=7.5cm]{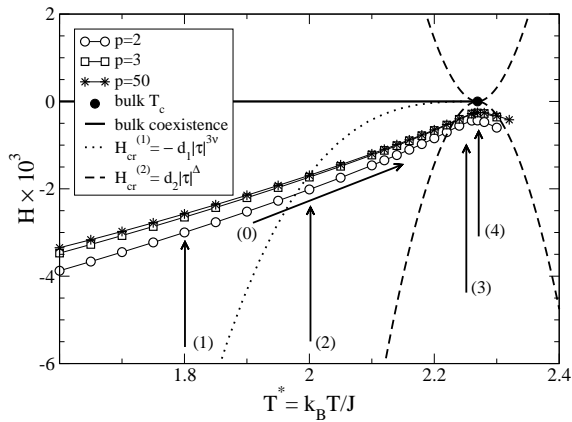}
\caption{Phase diagram for a  $d=2$ Ising strip subject to identical
boundary fields  $V^{ext}_{j}$ (see Eq.~(\ref{eq:3_2})) obtained by using
 DMRG  for a strip width  $L=340$ and the amplitude $h_1=0.8$
for  the boundary fields. The thick solid line indicates the 
bulk coexistence line. The lines interpolating the symbols represent the pseudo-phase coexistence
lines $H_{ca}(T,L;p)$ 
 for three different values of the exponent $p$ governing the 
algebraic decay 
of the boundary fields: open  circles correspond
to  $p=2$, open squares  correspond to $p=3$,
and  stars correspond  to  $p=50$.
Various crossover lines $H_{cr}^{(1)}$ and $H_{cr}^{(2)}$ with
$d_1=1=d_2=1$ and $\nu=1$ and $\Delta=15/8$ in $d=2$, as discussed in
 the main text,  are shown as well
as the thermodynamic paths $(0)-(4)$
along which  the adsorption has been calculated.
$(1)-(4)$ are various  isotherms and $(0)$ runs parallel to the pseudo-coexistence line.
Thus path (0) resembles (in a certain sense, see later) a  path of type
(I) in Fig.~\ref{fig:0} whereas  the paths $(1)-(4)$ do
correspond to a  path of type (II) in Fig.~\ref{fig:0}.}
\label{fig:1}
\end{figure}

In Fig.~\ref{fig:1} we show the phase diagram for the present model
calculated by using the DMRG method for a strip of  width  340
with $h_1=0.8$ and  three
choices of the parameter $p$ describing the decay of the boundary field: 
$p=50, 3$, and 2. The case   $p=50$ is expected to resemble the behavior 
corresponding   to  short-ranged surface forces. 
In this figure we  display the various crossover lines discussed in 
Sects.~\ref{sec:1} and \ref{sec:2} and the thermodynamic paths along which we have calculated
the adsorption  $\Gamma$.
The pseudo-phase coexistence line, the crossover lines, and the adsorption
have been determined for $2d$ Ising strips within the DMRG method.
\vspace{0.65cm}

The thick solid  line  in  Fig.~\ref{fig:1} indicates 
the bulk phase coexistence line $(H=0, T<T_c)$
terminating at the bulk critical point  $(H=0,T=T_c=[\ln(1+\sqrt 2)]^{-1}J/k_B\simeq 2.269 J/k_B)$. 
The  symbols (open circles for $p=2$, open  squares 
for $p=3$, and stars  for $p=50$) show the pseudo-phase coexistence.
 It turns out  that the pseudo-phase coexistence line
for long-ranged boundary fields is located slightly further 
away from the bulk coexistence line, especially at  lower temperatures,
 than the pseudo-phase coexistence line
for the short-ranged boundary fields ($p=50$).
These pseudo-lines   have been identified as those  positions $(H,T)$
in the phase diagram where  the total
magnetization of the strip vanishes, i.e., $\sum_{j=1}^Lm_{j}=0$
with  $m_{j}=\langle \sigma_{k,j} \rangle$.

At fixed $T$ (fixed $H$) at those positions the free energy of the strip exhibits 
a maximum as a function of $H$ ($T$) which is the rounded remnant of the nonanalyticity
(kink) of the free energy at the first-order phase transition
in the bulk ($L=\infty$) system. As can be seen in  Fig.~\ref{fig:1},
for $T<T_c$ $(T>T_c)$  the
line defined by the zeros of the total magnetization moves 
to less (more) negative  values of $H$ upon increasing
$T$. Determining the pseudocritical temperature $T_{c,L}$, where 
the pseudo-phase coexistence line ends, is difficult
because  in this quasi-one-dimensional system the critical point is not a sharp concept
and one has to examine various criteria in order to estimate $T_{c,L}$.
However, these different criteria, such as the
 maximum of the specific heat, or
the erosion of the jumps in the adsorption and in the solvation force
upon crossing the line  $H_{ca}(T,L;p)$,
provide somewhat different estimates for $T_{c,L}$. Determining 
 $T_{c,L}$ for different values of $p$ and $L$ is beyond the scope of the
present analysis. For the width $L=340$ and  $h_1=0.8$
we approximately obtain  $T_{c,L}\approx 2.2$; 
for more details concerning the short-ranged case
see  Ref.~\cite{maciolek:01}. 

\begin{figure}[h]
\centering
\includegraphics[width=7.5cm]{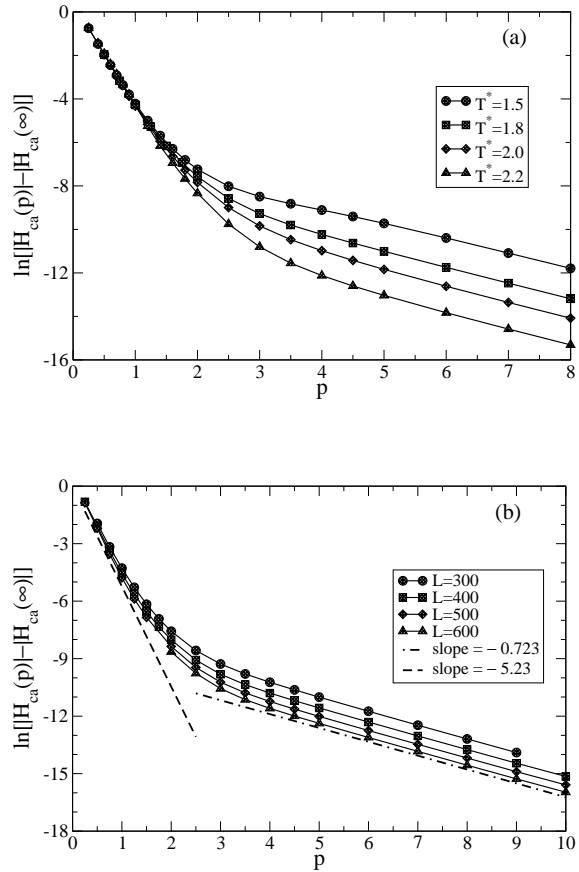}
\caption{The relative shift of the  pseudo-coexistence line  for Ising strips
as function of the range of the  boundary fields characterized by
the decay exponent  $p$ (Eq.~(\ref{eq:3_2})). The calculations are carried out by
using the DMRG method for  $h_1=0.8$ and (a) at fixed  $L=300$ and for four
temperatures  $T^*=k_BT/J$; (b) at fixed temperature ($k_BT/J=T^*=1.8$) and for
four thicknesses $L$ of the strip.}
\label{fig:2}
\end{figure}

We  have investigated  how  the location of the pseudo-phase coexistence 
line $H_{ca}(T,L;p)$ of capillary condensation changes as a function of $p$
for  fixed width $L$,  fixed amplitude  $h_1$ of the surface field, and fixed temperature $T$. 
In Fig.~\ref{fig:2} we plot the logarithm 
of the  difference  
$H_{ca}(T,L;p)-H_{ca}(T,L;\infty)$ as a function of $p$ calculated
for  $h_1=0.8$  and  (i) for several values of the temperature 
at fixed $L=300$ (Fig.~\ref{fig:2}(a))  as well as  (ii)
 for several values of $L$ at fixed $T=1.8 J/k_B$ (Fig.~\ref{fig:2}(b)). 
(Note that
$p=\infty$ corresponds to a pure surface contact field at $j =1, L$
(see Eq.~(\ref{eq:3_2})).)
From these  plots we can clearly distinguish two different regimes for 
the behavior of the
pseudo-phase coexistence  line, which are separated by a  crossover region 
occurring for
$1\lesssim p \lesssim 3$. In both of these two regimes the shift 
$H_{ca}(T,L;p)-H_{ca}(T,L;\infty)$ varies  exponentially  but
with  different decay constants.
Moreover,  for $p\lesssim 1 $ the shift relative to the
short-ranged pseudo-phase coexistence line does not   depend on the
temperature; the dependence on the strip width $L$ becomes,
for the  range of $L$ considered here, very weak and finally negligible for $p\lesssim 0.25$, i.e., 
$H_{ca}(T,L;p)-H_{ca}(T,L;\infty)\approx B_0\exp (-Ap)$ with $A\simeq 5.2$
and $B_0\simeq 1.5$.
For $p\gtrsim 3$ one has   $H_{ca}(T,L;p)-H_{ca}(T,L;\infty)\simeq  B(L,T)\exp (-Cp)$
with $C\simeq 0.723$. We note that, as expected, for $p$ fixed wider  strips and 
higher  temperatures give rise  to smaller shifts of the pseudo-phase coexistence line.
\vspace{0.65cm}

\begin{figure}[h]
\centering
\includegraphics[width=7.5cm]{rys03a.eps}
\end{figure}
\vspace{0.1cm}

\section{Determination of the  wetting temperature for different ranges of the boundary fields}
\label{sec:wtemp}

It is important to assure that the range of temperatures, for which
we perform our   calculations  of the adsorption, lies above 
the wetting temperature  $T_w(p,h_1)$.
 For $h_1=0.8$ and a short range of the  surface fields $(p=\infty)$ 
the wetting temperature  in the semi-infinite $d=2$
Ising system equals   $T_w(p=\infty,h_1=0.8)\simeq 1.41 J/k_B$
 and the wetting transition is second order.
 Adding  the  long-ranged tail to the boundary fields shifts 
the wetting temperature. Wetting transitions 
in $d=2$ in the presence of long-ranged substrate  potentials
have  been discussed by Kroll and Lipowsky \cite{kroll} within  the continuum
one-dimensional  solid-on-solid model, with the conclusion that 
 there is no wetting transition for fields which
decay more slowly than $1/j^3$. In this case their prediction is that 
the interface remains pinned to the wall at all finite temperatures.
This conclusion has been  supported  by providing  upper and lower bounds
 for the ground-state energy $E_0$ of the corresponding Schr\"odinger equation.
Within  this approach the existence of the bound-state
 solution of the Schr\" odinger equation ($E_0<0$) corresponds
to a  localized interface. When the bound state ceases to exists, $E_0\to 0$
signals  the critical wetting transition.
 The construction of upper and lower bounds (based on using an exponentially
decaying trial wave function) shows 
that the ground-state energy has a non-zero  value 
(i.e., there is no transition) for a finite
potential strength.
The  case $p=2$ has been studied
 by Privman and {\^S}vraki{\'c}
in Ref.~\cite{privman} also
within the SOS model; they  argued  that  for  attractive effective  potentials
decaying like $1/r$ the wetting transition is no longer sharp (becomes rounded).
 For the  $d=2$ Ising ferromagnet the wetting phenomena for the case of the
marginal range  $p=3$ have been studied
within the DMRG approach \cite{drzew} and by MC
 simulations \cite{albano,devirgiliis}.
These studies have shown that the presence of long-ranged tails 
in the boundary fields decreases the wetting temperature with respect to the
case of the short-ranged surface fields and that 
  $T_w(p=3,h_1=0.8)\simeq 0.75J/k_B $.
Apart from Refs.~\cite{kroll,privman}, 
the case  $p<3$ has not been studied so far.

In order to check the predictions by Kroll and Lipowsky,
  we investigate whether in the $d=2$ Ising
 ferromagnet with $p=2$ a  wetting transition exists. 
In the  finite systems we are studying,  below the bulk
 critical temperature  the wetting 
transition as a function of  $h_1$
for any range $p$  of the boundary fields can be inferred
from  the so-called  (weakly rounded in $d=2$)
interface localization-delocalization (ILD) transition 
\cite{binder:08,parryevans} which
occurs in strips with antisymmetric  boundary fields, i.e., for
$V_{j}^{ext}=V_{j}^s-V_{L+1-j}^s$. This transition is the precursor
of a wetting phase transition that occurs in the limit  of infinite film
thickness $(L\to\infty)$ at the critical curve  $T_w(p,h_1)$. For $T<T_w(p,h_1)(T>T_w(p,h_1))$
such an interface is bound to (unbound from) the walls \cite{binder:08,parryevans}.

First, we note  that $T_{w}(h_1\to 0,p)\to T_c$ for any $p$ (see Eq.~(\ref{eq:3_2})).
On the other hand  $h_{1w}(T=0,p)$, which is the solution of the implicit equation 
$T_w(h_1,p)=0$ and denotes the critical surface field 
strength beyond which the system is wet even at $T=0$, shifts towards lower 
values upon decreasing  $p$. 
There is no reason to expect a non-monotonic behavior of $T_w(h_1,p)$.
Therefore,  the gross features  of the shape of the 
wetting transition line $T_w(h_1,p)$ for an arbitrary $p> 1$ 
can be inferred from localizing  the  position
of the wetting transition $h_{1w}(T=0,p)$ in the ground state.
The ground-state energy of the system (in units of $J$) 
can be found directly from the Hamiltonian with a vanishing bulk field. 
Because the system is translationally invariant  along a strip,
it is sufficient to consider only the configurations of a single column.

In a {\it p}artial {\it w}etting regime with antisymmetric surface fields
there are only two coexisting states (all spins up or all spins down)
with the energy (per the number $M$ of columns)
\begin{widetext}
\begin{eqnarray}
E_{pw}^{\pm} & = & -J(L-1)\mp h_{1}J \left( 1-\frac{1}{L^p}+\frac{1}{2^p}-\frac{1}{(L-1)^p}+\ldots \right. 
\mbox{}+\left.\frac{1}{(L-1)^p}-\frac{1}{2^p}+\frac{1}{L^p}-1 \right) \nonumber \\
&=&-J(L-1).
\end{eqnarray}
\end{widetext}
In a wet regime at least one interface between spin up and spin down configurations has to be present.
For even  $L$  the lowest-energy configuration 
is that for the state with the interface located  in the middle of the strip.
For  $p \to \infty$ (short-ranged case) $L-1$ degenerate states emerge with a  single interface
positioned at any of the rows but the ones closest to the
surfaces. In general, the energy of a single column with an interface in the middle  of a strip (spin up for
$j=1,\ldots, L/2$ and spin down for $j=L/2+1,\ldots, L$) is
\begin{widetext}
\begin{eqnarray}
\label{eq:gs_2}
E_{wet} & = & -J(L-3)-h_{1}J \left( 1-\frac{1}{L^p}+\frac{1}{2^p}-\frac{1}{(L-1)^p}+\ldots \right. 
  \mbox{}+\frac{1}{(L/2)^p}-\frac{1}{(L/2+1)^p} \nonumber \\
&-&\frac{1}{(L/2+1)^p}+\frac{1}{(L/2)^p}+\ldots
\mbox{}-\left.\frac{1}{(L-1)^p}+\frac{1}{2^p}-\frac{1}{L^p}+1\right).
\end{eqnarray}
\end{widetext}
In the ground state and at the ILD transition the  energies $E_{pw}$ and $E_{wet}$ are equal.
For a particular  $L$ this  determines the magnitude  $h_{1w}^{ILD}(L)$:
\begin{widetext}
\begin{eqnarray}
\label{eq:gs_3}
h_{1w}^{ILD}(L)& = & 1/\left(1+\frac{1}{2^p}+\frac{1}{3^p}+\ldots+\frac{1}{(L/2)^p}-\frac{1}{(L/2+1)^p}-\ldots-\frac{1}{(L-1)^p}-\frac{1}{L^p} \right)\nonumber \\
& = &
\left\{ \sum_{n=0}^{\infty}\frac{1}{n^p}-2\sum_{n=L/2+1}^{\infty}\frac{1}{n^p}+\sum_{n=L+1}^{\infty}\frac{1}{n^p} \right\}^{-1}.
\end{eqnarray}
\end{widetext}
In order to find the critical  wetting field $h_{1w}(L\to\infty)$  
we take the limit $L\to\infty$. According to the second line in 
Eq.~(\ref{eq:gs_3}) this leads to
\begin{equation}
\label{eq:gs_4}
h_{1w}=1/\sum_{n=0}^{\infty}\frac{1}{n^p}=1/\zeta(p)
\end{equation}
where $\zeta(p)$ is the Riemann zeta function. 
Its values are known analytically only for certain
even values: $\zeta(p=2)=\pi^2/6$, $\zeta(p=4)=\pi^4/90$ which 
 gives $h_{1w}(p=2)\approx 0.6079$ and 
$h_{1w}(p=4)\approx 0.9239$. 
Other values can be found in  tables of special functions, e.g.,
$\zeta(p=3)\approx 1.2021$ giving $h_{1w}(p=3)\approx 0.8319$.
In the short-ranged limit ($p \to \infty$) the Riemann zeta 
function approaches  $1$, confirming  Abraham's  solution at $T=0$ \cite{abraham}.
Even  more interesting is the opposite limit $\zeta(p \to 1) \to \infty$ which 
 results in $h_{1w}(p\to 1) \to 0$. This  means that at $T=0$
the wetting transition does not exist for $p \leq 1$.
\vspace{0.65cm}

\begin{figure}[hbt]
\centering
\includegraphics[width=8.0cm]{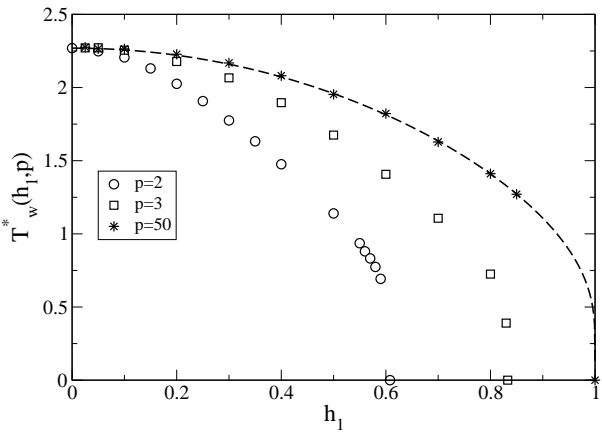}
\caption{Phase diagram at bulk coexistence  $H=0$ for continuous
wetting transitions in the $d=2$
Ising ferromagnet for (Eq.~(\ref{eq:3_2})) $p=2$, $p=3$, and $p=50$
(quasi-short-ranged boundary field) obtained within the DMRG approach.
For $p>1$ the values in the limit $T\to 0$ are  known exactly for semi-infinite systems.
The dashed line is the analytically known  exact result for $p=\infty$
and for semi-infinite systems ~\cite{abraham}. $T^*_w=k_BT_w/J$.}
\label{fig:3}
\end{figure}

In order to determine the location of a quasi-ILD transition between $T=0$
and $T=T_c$ for $p<3$ various criteria can be
applied \cite{drzew}. Here we have adopted  the approach
involving  the  magnetic susceptibility $\chi$.
The singularity (or a maximum) of the magnetic susceptibility $\chi$ is one
of the most useful criteria for the localization of a phase transition (or of
a pseudo-phase transition for a finite system). The magnetic susceptibility can
be calculated as the second derivative of the free energy 
$f$ with respect to the bulk magnetic field $H$. This method  is very convenient for 
the DMRG approach because the latter provides the free energy with a very high accuracy.
Nevertheless the present  case is somewhat  special, because we want to
determine $T_w(p,h_1)$  at $H=0$ (see Fig.~\ref{fig:3}), 
where in the partial wetting regime, i.e., for $T<T_w$,
there is a first-order bulk transition. In the  thermodynamic limit
there is coexistence of phases with opposite magnetizations.
Thus there is a discontinuity of the first derivative of the free energy $f$ (a jump
of the magnetization $m=-\partial f/\partial H$) upon changing the sign
of the bulk magnetic field. Accordingly, in order to calculate $\chi$ there, 
one has to calculate the derivatives for small nonzero bulk fields and then to consider 
the limit  $H\to 0$. In the complete wetting  regime, i.e., $T>T_w$, or equivalently  above the
ILD transition, the finite system exhibits
only a single phase with  an interface meandering  freely between the walls so that there
is no discontinuity of the free energy derivatives upon crossing $H=0$.
For numerical calculations such as the ones within  the DMRG method
the necessity of performing an extra limiting procedure ($H \to 0$ in this case) is cumbersome. 
Therefore, instead of $\chi$,  we have focused on another quantity $\chi_{0}$,
 which  also corresponds to the second derivative
of the free energy at fixed $T$ and $h_1$, but is calculated numerically  
in a symmetrical way with
respect to  $H=0$  by taking  the free energy values at  five equidistance
points: $-2\Delta H$, $-\Delta H$, $0$, $\Delta H$ or $2\Delta H$; we typically
used $\Delta H = 10^{-5}$. Because our calculations are always carried out for
finite $L$, there is no discontinuity of the magnetization 
in the partial wetting  regime.
These discontinuities  are replaced by functions which are
rounded but  steeply varying at $H=0$.
   In order to determine the ILD transition we have scanned the phase diagram at fixed $h_1$.
The higher the temperature, the less steeply  the  magnetizations vary
and the values of their
derivative $\chi_{0}$ are smaller. Above the wetting temperature, 
where there is no  discontinuity,
  $\chi_{0}(H=0,T)$ saturates for increasing  $T$; here $\chi_{0}$ is
equivalent to $\chi$. Therefore, at fixed $L$, the ILD transition can be
 identified 
by the maximal slope of  $\chi_{0}$ or the minimum of its derivative with respect to
temperature. Although all derivatives have been performed numerically,
the high accuracy of the DMRG method  guarantees very precise results.
\vspace{0.65cm}

\begin{figure}[h]
\centering
\includegraphics[width=7.5cm]{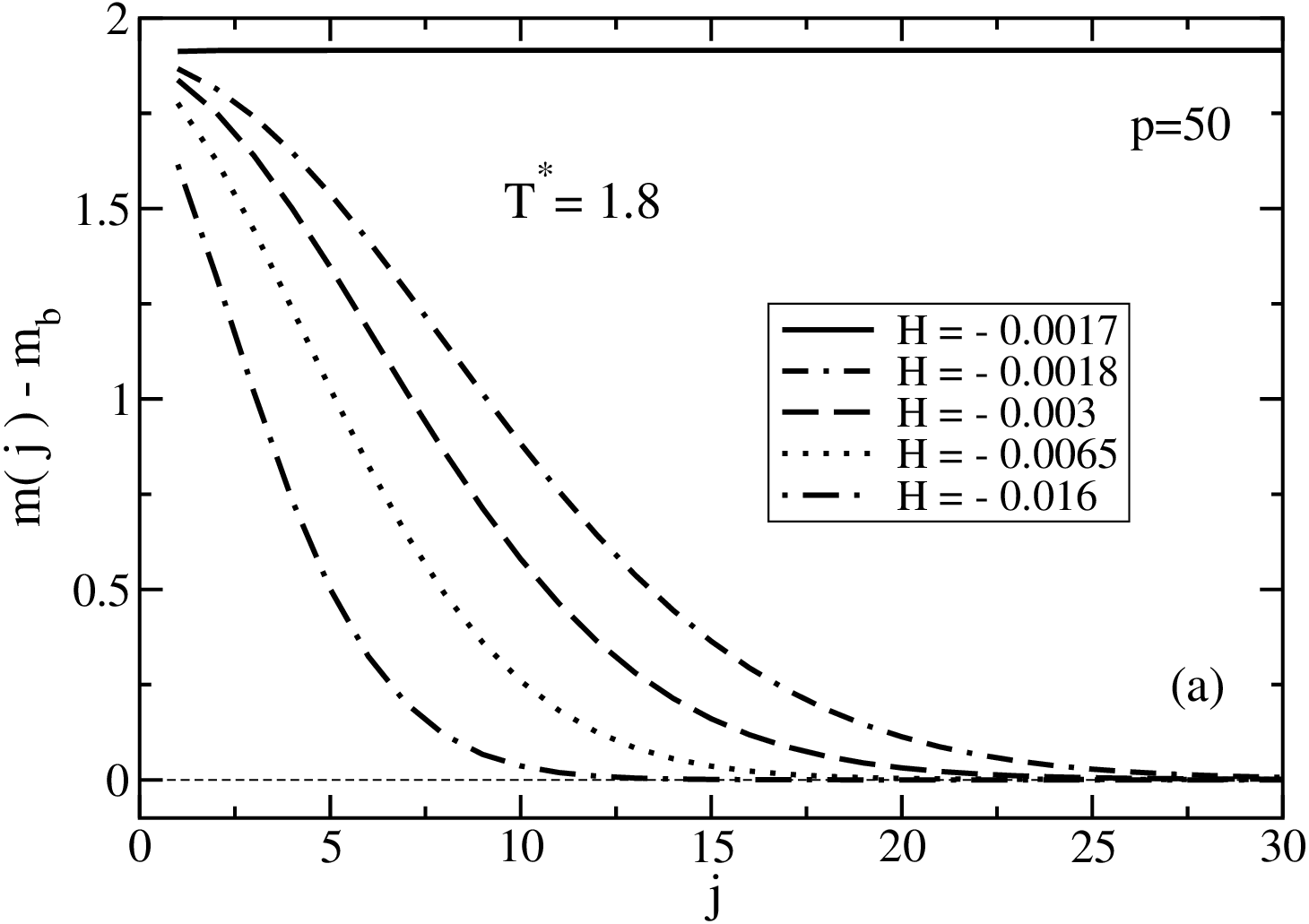}
\end{figure}

\begin{figure}[h]
\centering
\includegraphics[width=7.5cm]{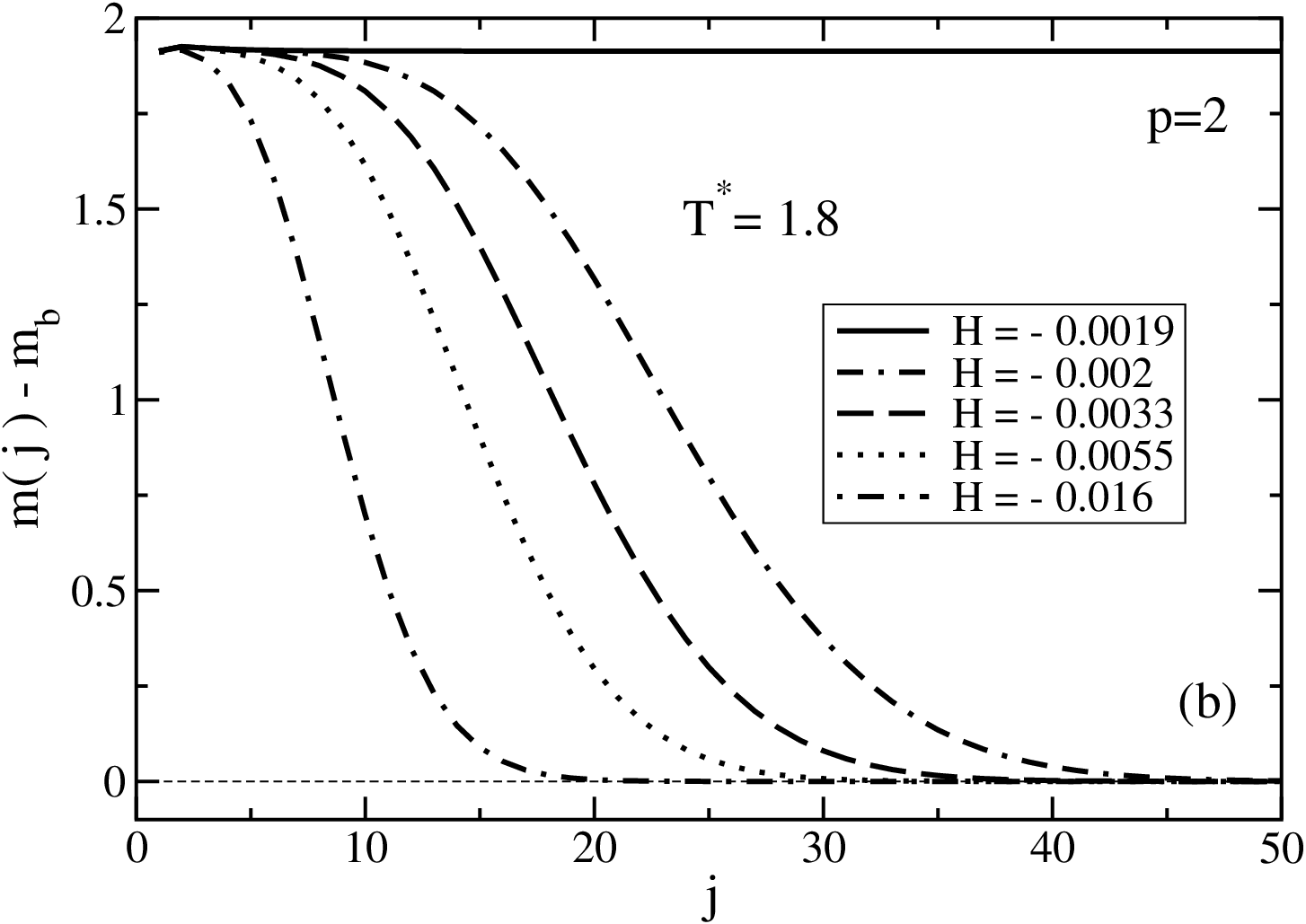}
\caption{Magnetization profiles, relative to their bulk values
$m_b(T,H)<0$, near one wall ($j >1$) along the isotherm $T^*=1.8$ 
(see path (1) in Fig.~\ref{fig:1}), calculated within the DMRG method for $d=2$ Ising strips of
width $L=500$, for (a) short-ranged ($p=50$) and (b) for
long-ranged ($p=2$) boundary fields of strength $h_1=0.8$.
Along this isotherm the pseudo-capillary condensation transition occurs
at $H_{ca}\simeq -0.00172$  for $p=50$ and at $H_{ca}\simeq -0.00195(4)$ for $p=2$.
Thus both in (a) and (b) the full line  for $H=-0.0017$ and $H=-0.0019$, respectively,
corresponds to the capillary filled state with $m(j)-m_b(T,H)\simeq 2|m_b(T,H)|$.}
\label{fig:4}
\end{figure}

Finally we have extrapolated $T_w(p,h_1;L)$ to the limit $L\to \infty$ in order to obtain the wetting
temperatures $T_w(p,h_1)$. The obtained wetting phase diagram in the $(T,h_1)$ plane
is shown in Fig.~\ref{fig:3} for $p=2, 3$, and  $p=50$.
The dashed line is the exact result by Abraham for $p=\infty$ \cite{abraham}.
The close agreement of our data for $p=50$ with this exact result generates
confidence in our numerical procedures.
Our results show that contrary to the claim by Kroll and Lipowsky \cite{kroll} 
also for $p=2$ there are wetting transitions and that the wetting temperature 
of these continuous transitions can be well localized as a function of $h_1$.
In view of the calculations of the adsorption, for our choice $h_1=0.8$ 
the  surface is wetted for all temperatures. As already mentioned above, the claim by Kroll 
and Lipowsky is based on using a continuum planar 
solid-on-solid (SOS) approximation. SOS models ignore bulk-like fluctuations
such as bubbles and the formation of "overhangs" of the line separating 
oppositely magnetized regions, while focusing on describing  the long-wavelength
behavior of the interface, expected to play the crucial role in wetting phenomena. 
In the present case of the $d=2$ Ising model the bulk fluctuations 
are particularly strong giving rise to a very diffuse interface.
For the case of  $p=2$ they might provide the mechanism 
for the unbinding of the interface from the wall, even though 
the long-wavelength interfacial fluctuations are strongly 
supressed by this very long-ranged boundary field. 

\section{Numerical results and discussion}
\label{sec:res}   

Using the DMRG method we have performed calculations of the
magnetization  profiles $m(j)$ from which we have obtained  the
thicknesses  $\ell_0$ of the wetting layer and the adsorption $\Gamma$.
The analysis of the  shapes of the profiles  $m(j)$  and of 
the scaling properties of
 $\ell_0$ gives a better 
understanding of the behavior of $\Gamma$ along different
thermodynamic paths. We recall, that the predictions
 for the scaling behavior of the adsorption follows from the
scaling behavior of $\ell_0$.

In order  to infer possible power laws  governing
the behavior of the thickness of the wetting layer and of
the adsorption   along different paths   we have
calculated  {\it local exponents} of the quantities of  interest
as a function of $H$ or $\tau$.
They are defined as
\begin{equation}
z_i = \Bigl\lvert \frac{\ln Q (i+1) - \ln Q (i)}{\ln x_{i+1} - \ln x_{i}}\Bigr\rvert \,\, ,
\label{zN}
\end{equation}
which is the discrete derivative of  data $Q$ as a function 
of $x$ in a log-log  plot;  here $Q=\ell_0$ or $\Gamma$ and $x=H$ or $\tau$.
Such a quantity probes the local slope
at a given value $x_i$ of $x$ at a point $i$ along 
the path considered. It provides a better estimate
 of the leading exponent than a log-log plot itself. We have chosen the
convention that if $Q$ decays
algebraically as a function of $x$ then $Q\sim x^{-z}$ with $z>0$.
The high quality of the DMRG data allows us to reliably 
carry out this numerical differentiation.

\subsection{Isotherms}
\label{subsec:iso}

\subsubsection{Magnetization profiles}
\label{subsubsec:mp}

\begin{figure}[h]
\centering
\includegraphics[width=7.5cm]{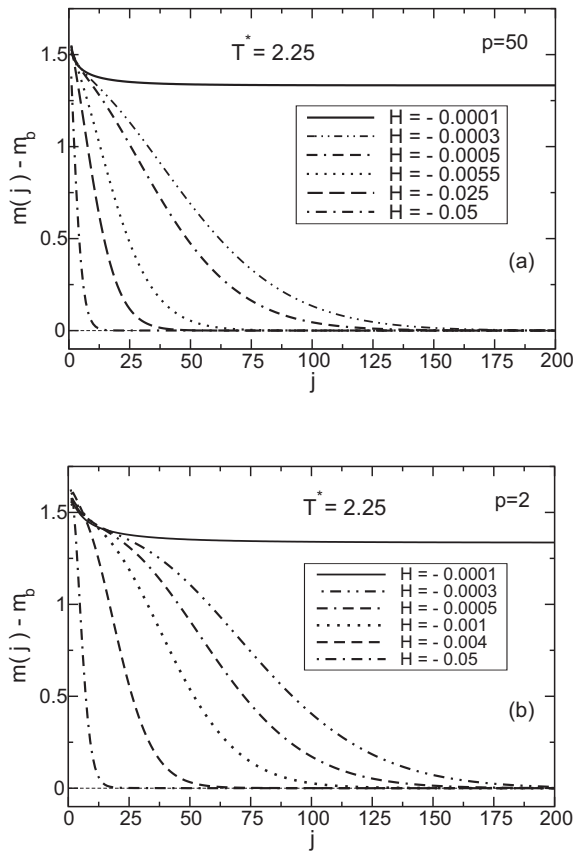}
\caption{Magnetization profiles, relative to their bulk values
$m_b(T,H)<0$, near one wall along the isotherm $T^*=2.25$ (see path (3) in Fig.~\ref{fig:1}),
calculated within the DMRG method for  $d=2$ Ising strips of  width $L=500$,
(a) for short-ranged ($p=50$) and (b) for long-ranged ($p=2$) boundary fields of strength $h_1=0.8$.
Along this isotherm the pseudo-capillary condensation transition occurs
at $H_{ca}\simeq -0.000191$  for  $p=50$ and at $H_{ca}\simeq -0.00026(4)$ for  $p=2$.
Thus both in (a) and (b) the full line  for $H=-0.0001$ corresponds
to the capillary filled state with $m(j)-m_b(T,H)\simeq 2|m_b (T,H)|$.}
\label{fig:5}
\end{figure}

In Figs.~\ref{fig:4} and \ref{fig:5}  we show a selection of magnetization
profiles calculated for a strip width $L=500$ with   $p=50$
 and 2 along the  two isotherms
$T^*=1.8$ and  $T^*=2.25$ indicated in Fig.~\ref{fig:1} as path (1) and (3), respectively.

\begin{figure}[h]
\centering
\includegraphics[width=7.5cm]{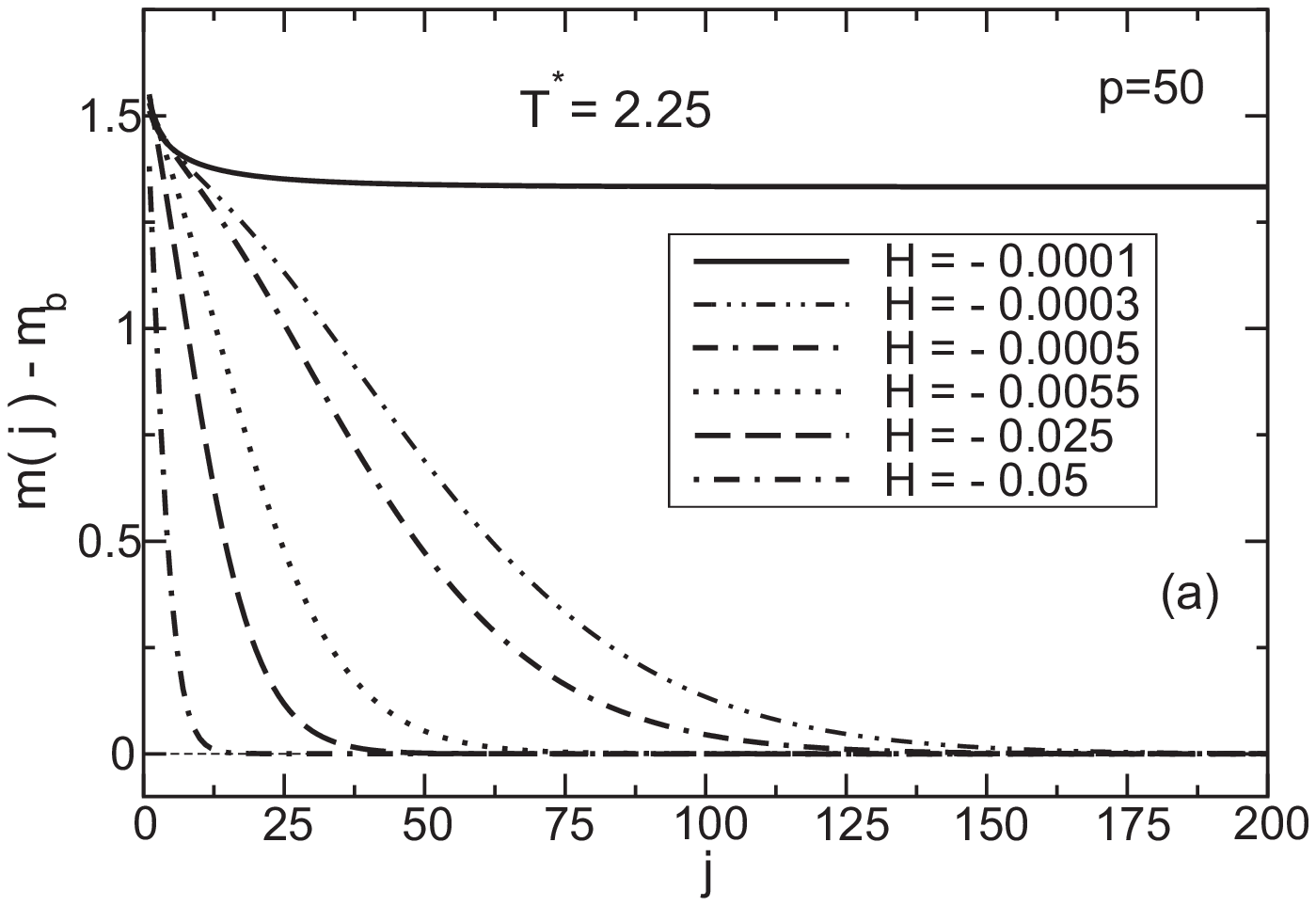}
\end{figure}

From these plots one can see that within the accessible range of
values $H<H_{ca}(T,L;p)$  the shapes of
the profiles are {\it not} slab-like, even though the system size is very big
so that $|H_{ca}(T,L)|$ is expected to be sufficiently small.
 They are characterized by a pronounced spatial  variation at the
emerging interface between the spin up and the spin down phase,
 where the profile varies quasi  linearly
 with the distance from the wall, and by extended  tails.
For the short-ranged  case ($p=50$) the profiles along the low-temperature
isotherm ($T^*=1.8$) exhibit a rapid decay
 to their bulk values $m_b \equiv m_b(T,H)$ ($<0$ for $H<0$) with the bulk correlation length
$\xi(T,H)= |\tau|^{-\nu}\Xi(H|\tau|^{\Delta})$
(see Eqs.~(\ref{eq:6})-(\ref{eq:ximu})) as decay length
which increases upon approaching the pseudo-phase coexistence.
Only very close to the pseudo-phase coexistence line $H_{ca}(T,L;p)$
the emerging shape of the profile to a certain extent  resembles that
characteristic of the free  interface between the spin up and the spin
down bulk phases with a broad interfacial region
and a rather narrow  region where the magnetization stays
somewhat  close to the value
$-m_b>0$ characteristic for the wetting
phase corresponding to $H=0^+$;  this latter region
thickens as $H\to H_{ca}(T,L;p)$.

For the long-ranged forces ($p=2$) the  profiles
along the low-temperature isotherm  ($T^*=1.8$)  
exhibit an  interface-like shape  already away from
the pseudo-phase coexistence line $H_{ca}(T,L;p)$. 
The interfacial part of these profiles is  much  
more pronounced than for the short-ranged case ($p=50$) and 
their  decay  is much  slower.
The exponential decay of the profiles
to the value $m_b(T,H)$ ultimately crosses over to the power law
decay  $\sim j^{-2}$. In general, algebraically decaying inter-particle
and surface fields are known to generate algebraically decaying order parameter
profiles \cite{SN}.

For  both $p=50$ and $p=2$  the profiles corresponding to $H\to 0$ along
the high-temperature isotherm  ($T^*=2.25$) decay much  slower as compared to the low-temperature 
isotherm ($T^*=1.8$) and their shapes become quantitatively different (see  Fig.~\ref{fig:5}).
The magnetization in the first few layers near the walls  decreases more rapidly so that there 
the profiles acquire a positive curvature. Further away from
the wall there is an inflection point and the  narrow pleateau occurring 
near the boundaries for the low-temperature isotherm (see Fig.~\ref{fig:4}(b))
disappears. The interfacial region becomes  very broad.

\subsubsection{Thickness of the wetting layer}
\label{subsubsec:wl}

In order to infer equilibrium  thicknesses $\ell_0$ 
 of the wetting layers from
the magnetization profiles we have chosen the criterion of the
change of sign of the curvature of the profile, i.e., 
we have assigned a thickness to a  wetting layer  which
corresponds  to the distance $j=\ell_0$ at which the profile 
exhibits its  inflection point. 
Adopting other criteria for determining 
the wetting film thickness, e.g., the thickness which corresponds
 to the distance $j=j_0$ 
at which the magnetization 
vanishes, $m(j_0)=0$, leads, as it must be, to the same conclusions about physical observables.

We have studied  the behavior of $\ell_0$ along several isotherms.
Representative data are shown  and discussed below. They correspond
to the isotherms $T^*= 1.8,  2.25$ 
(see the paths (1) and (3), respectively,  in Fig.~\ref{fig:1}) and $T=T_c$ (see path (4)
in Fig.~\ref{fig:1}) and have been obtained 
for several strip widths and ranges $p=50,5, 4, 3, 2, 1.5$ of the boundary fields.
(The data for the isotherms $T^*=1.6$, 1.9, 2,  and 2.2 are not shown.)
We have chosen  $h_1=0.8$ for which $T^*_w(p=\infty)\simeq 1.41$. 
All considered isotherms lie above the corresponding wetting temperatures  $T_w(p)$.

\begin{figure}[h]
\centering
\includegraphics[width=7.5cm]{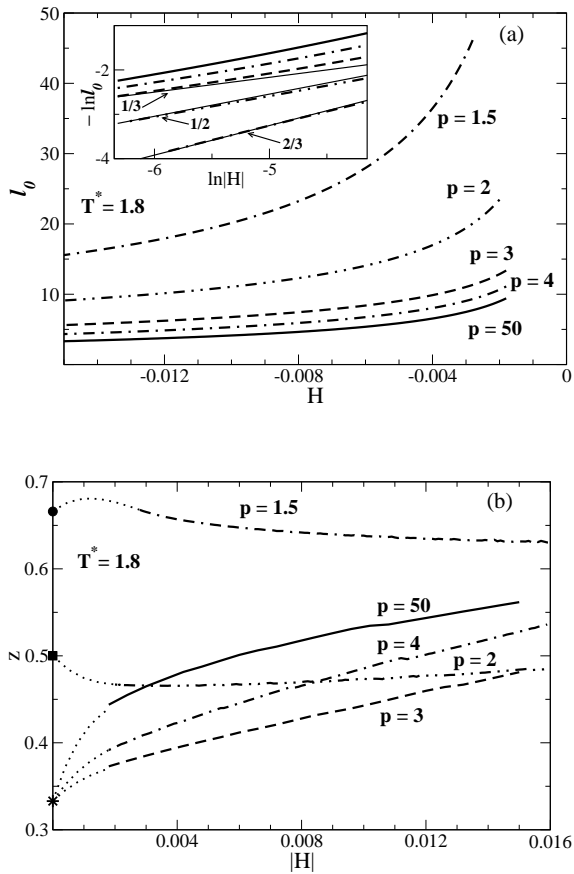}
\caption{ (a) Equilibrium wetting layer thickness $\ell_0$
in units of the lattice constant $a$ as a function of  the bulk magnetic field
$H$  calculated along the isotherm $T^*=1.8$
 for $L=500$, $h_1=0.8$ and for the  ranges $p=1.5, 2, 3, 4$, and  50 of the
boundary fields.
A rapid increase of $\ell_0$ upon approaching the pseudo-phase  coexistence
line $H_{ca}(T,L;p)$ supports  the existence
of the wetting transition for $p=2$ and 1.5. The values of $H_{ca}(T,L;p)$
are given by the points at which the various lines end on the right.
The inset shows  a log-log plot of these data. The thin solid lines indicate slopes
(from top to bottom): 2/3, 1/2, and 1/3.
(b) The local exponents $z$ for $\ell_0(H)$ (see Eq.~(\ref{zN})).
For $H\to 0$, without the occurrence  of capillary condensation, i.e., for
$L\to \infty$, the expected exponents are $\beta_s^{co}=1/3$ $(\star)$ for
$p=3, 4$, and 50, 1/2 $(\small{\blacksquare})$ for $p=2$, and 2/3 $(\bullet)$ for $p=1.5$.
The dotted lines indicate the expected extrapolations $H\to 0$ for $L\to \infty$.
The log-log plot supports these conclusions. We note that the local exponents do not always follow
from naive visual impressions. For example, $z(p=50)>z(p=2,3,4)$, although in (a) the curves of
$\ell_0(p=2,3,4)$ seem to increase somewhat stronger than the one for $\ell_0(p=50)$.}
\label{fig:6}
\end{figure}

Along  the isotherm $T^*=1.6$ the wetting layers  are so  thin
that it is difficult to assign a suitable thickness $\ell_0$, especially for $p=50$.
Along the isotherm at the  higher temperature $T^*=1.8$ (path (1) in Fig.~\ref{fig:1}) this 
assignment is much clearer and the results for  $\ell_0(H)$  and its local exponents 
are shown in  Fig.~\ref{fig:6}. They were obtained  for  strips of  width $L=500$ 
with   $h_1=0.8$ and for various ranges $p$ of the boundary fields.

\vspace{0.65cm}
\begin{figure}[h]
\centering
\includegraphics[width=7.5cm]{rys07a.eps}
\end{figure}
\vspace{0.1cm}

\begin{figure}[h]
\centering
\includegraphics[width=7.5cm]{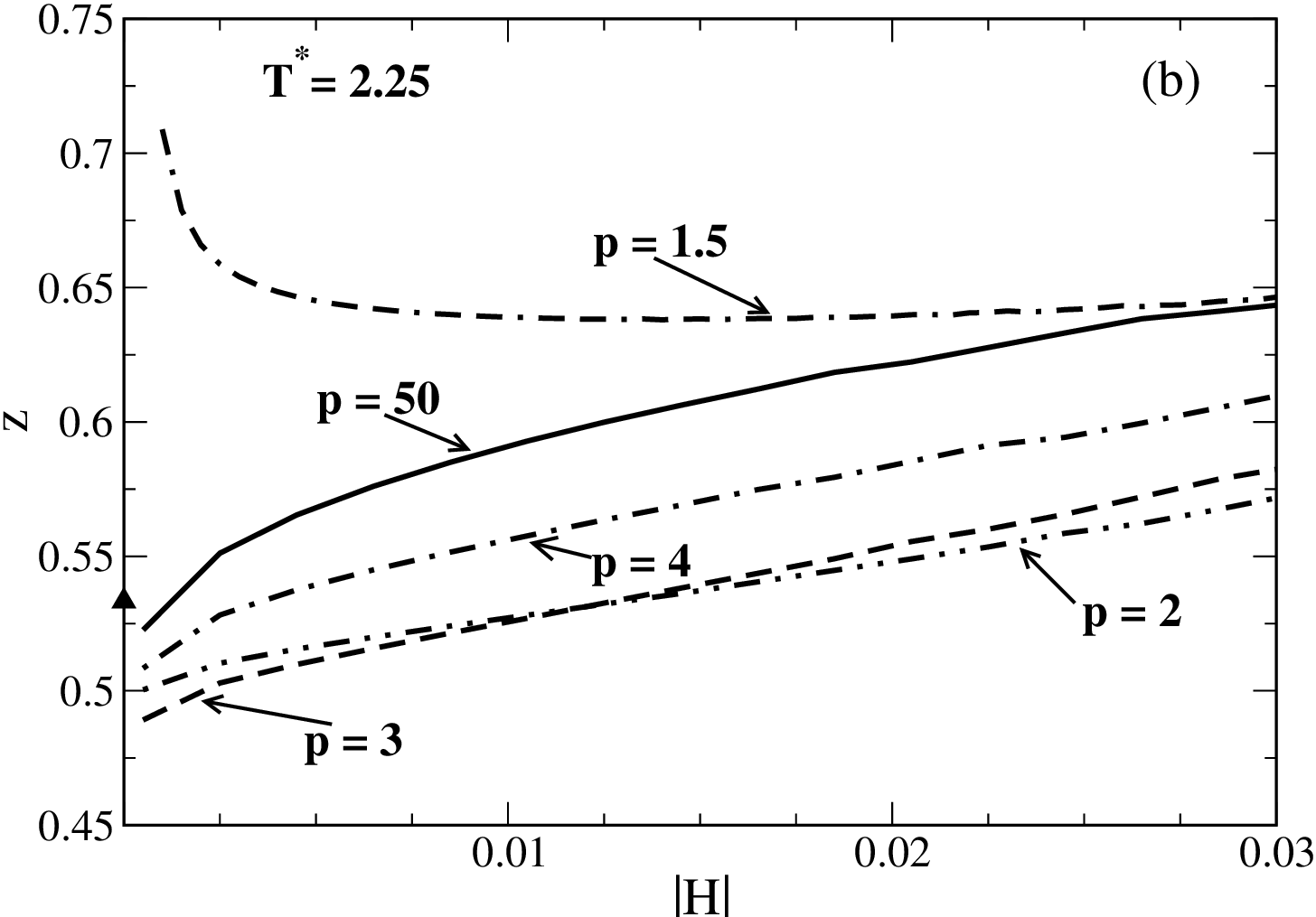}
\caption{ (a) Wetting layer thickness $\ell_0$
in units of the lattice constant $a$ as a function of the bulk magnetic field
$H$ calculated along the isotherm $T^*=2.25$ (path (3) in Fig.~\ref{fig:1})
for $L=500$, $h_1=0.8$, and for the  ranges $p=1.5, 2, 3, 4$, and  50 of the
boundary fields. On the right the curves end at the corresponding capillary
condensation point $H_{ca}(T,L;p)$. The inset shows  a log-log plot of these data. The thin 
solid line indicates the slope 8/15. (b) The local exponents $z$ for $\ell_0(H)$ (see Eq.~(\ref{zN})). 
For $p\ge 2$ and $H\to 0$ they tend towards 0.48, $\ldots $, 0.52 whereas the expected
value 8/15$ (\blacktriangle)$ is somewhat higher. According to Fig.~\ref{fig:1}, along this 
isotherm capillary condensation prevents one to reach the ultimate complete wetting regime.
For $p=1.5$ the behaviors of $\ell_0(H)$ and $z(H\to 0)$ are
distinctly different from those for $p\ge 2$.}
\label{fig:7}
\end{figure}

\begin{figure}[h]
\centering
\includegraphics[width=7.5cm]{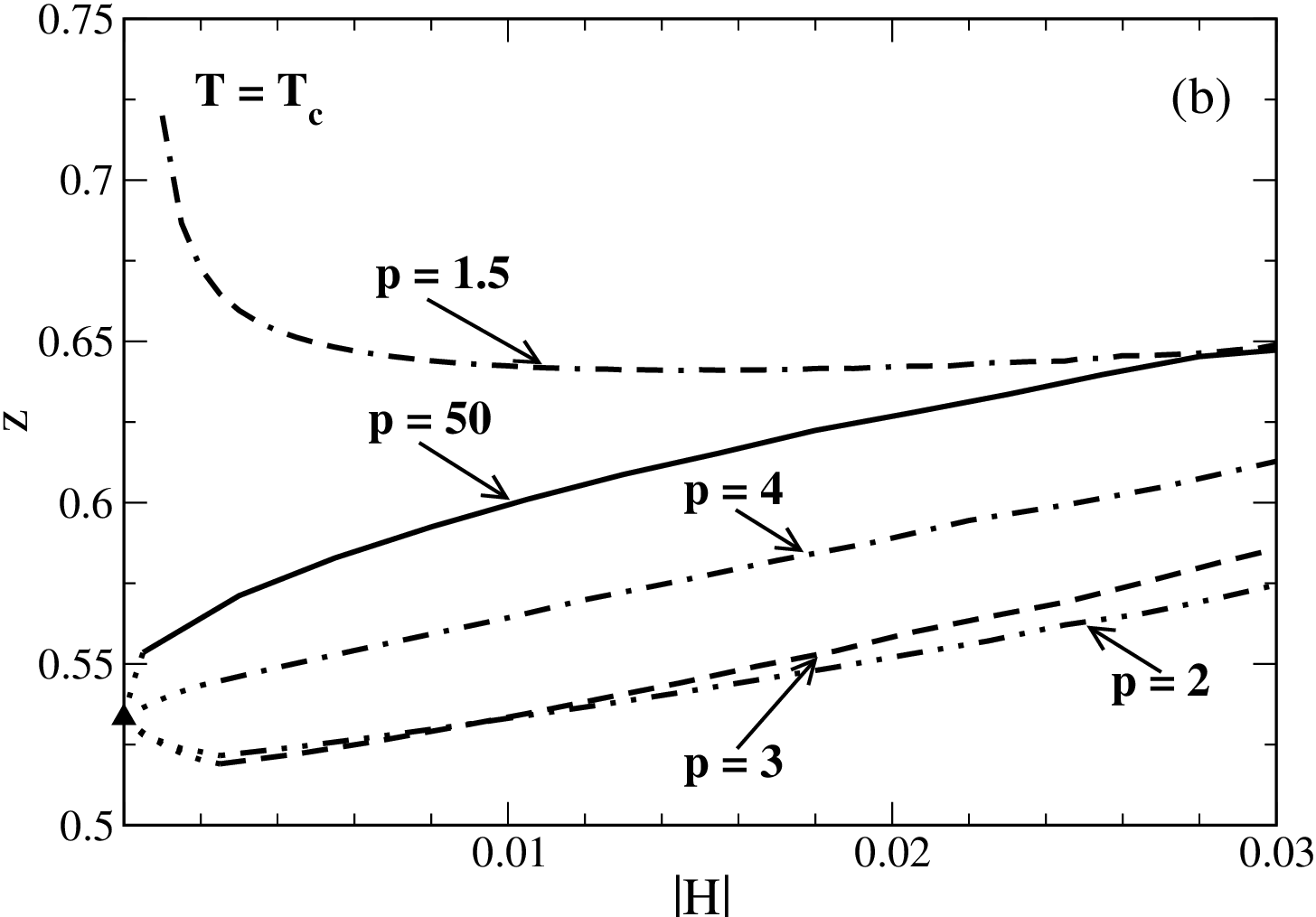}
\caption{ (a) Wetting layer thickness $\ell_0$ in units of the lattice constant $a$ as a function 
of the bulk magnetic field $H$ calculated along the critical isotherm $T^*=T^*_c=2.269$
(path (4) in Fig.~\ref{fig:1}) for $L=500$, $h_1=0.8$, and for the  ranges $p=1.5, 2, 3, 4$, 
and $50$ of the boundary fields. On the right the curves end at the corresponding capillary
condensation point $H_{ca}(T,L;p)$. The inset shows a log-log plot of these data.
The thin solid line indicates the slope 8/15. (b) The local exponents $z$ for $\ell_0(H)$ 
(see Eq.~(\ref{zN})). For $H\to 0$, without the occurrence of capillary condensation, i.e., for
$L\to \infty$, the expected exponent is $\nu/\Delta=8/15$  $(\blacktriangle)$.
The dotted lines indicate the expected extrapolation for $z(p\ge2,H\to 0$) with $L\to \infty$.
For $p=1.5$ the behaviors  of $\ell_0(H)$ and $z(H\to 0)$  are
distinctly different from those for $p\ge 2$.}
\label{fig:8}
\end{figure}

\begin{figure}[h]
\centering
\includegraphics[width=7.5cm]{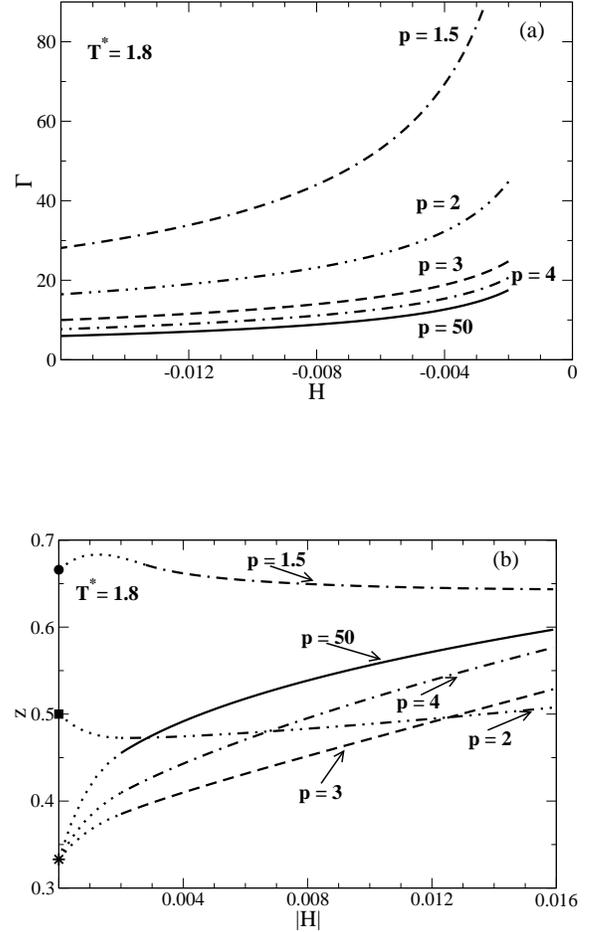}
\caption{ Adsorption $\Gamma(H)$ (a) and the corresponding  local exponents $z$ (b) along
the isotherm  $T^*=1.8$ (path (1) in Fig.~\ref{fig:1})
calculated within the DMRG approach for a $d=2$ Ising strip
of  width $L=500$ subject to symmetric  boundary fields of
strength  $h_1=0.8$ and various ranges $p$ corresponding to
power law decays $\sim j^{-p}$. For small $|H|$ all curves end at $H=H_{ca}(T,L;p)$
where capillary condensation sets in.
Dotted  lines indicate the expected extrapolations to $H=0$ for semi-infinite
systems:  $\beta_s^{co}=1/3$ $(\star)$ for
$p\ge 3$,  $\beta_s^{co}=1/2$ $(\small{\blacksquare})$  for $p=2$,  and
$\beta_s^{co}=2/3$ $(\bullet)$ for $p=1.5$.}
\label{fig:9}
\end{figure}

Path (1) lies entirely within the region between the bulk coexistence curve and the crossover boundary
 $H^{(1)}_{cr}= -d_1|\tau|^{3\nu}$ with $\nu =1$ for the $d=2$ Ising model, i.e., within the complete 
wetting regime (see Fig.~\ref{fig:1}). Indeed, for all values of $p$ we observe the increase of 
$\ell_0$ upon approaching the pseudo-phase coexistence line $H_{ca}(T,L;p)$. 
The lines in Figs.~\ref{fig:6}-\ref{fig:11} end on the right 
when they reach the pseudo-phase coexistence  line $H_{ca}(T,L;p)$ (see Fig.~\ref{fig:1}). 
 Apart from this size dependence of the $H$--window with access for 
complete wetting, for  $L=500$ and  300 (data for $L=300$ are not shown) 
 the finite-size effects are negligible.
As expected from the shape of the profiles
 (see Figs.~\ref{fig:4} and \ref{fig:5}) for the same value of $H$
the wetting layer is much 
thicker  for small
values of the exponent $p$ than for short-ranged  surface fields ($p=50$).
The rate of  increase of $\ell_0$, as characterized by the
behavior of the local exponents $z$, depends  on the range of the
boundary field  $p$ and   varies along the isotherm.
For $p=1.5$ the rate of  increase is always largest, i.e., 
$z$  remains  the largest   along the  entire isotherm
 (see  Fig.~\ref{fig:6}(b)). 
For  $p=50, 4$, and 3 the local
exponent $z$ decreases as $|H|\to 0$, i.e., 
 the initial increase  of the wetting layer 
 is larger than the one observed  very close to $H=0$ for which
 the plateau in the profile near the wall corresponding
to the wetting phase at  $H=0^+$ starts to form.
For $p=2$ the  very long-ranged nature of the boundary field 
enforces the occurrence  of 
 a narrow plateau in the profile near the wall even  far away from 
$H=0$. Upon decreasing $|H|$ the width of 
this plateau increases  and the emerging interfacial
 region broadens giving
rise to the increase of $\ell_0$ which is  initially  smaller  than 
for $p=50$. Only
very close to  $H=0$ this trend
reverses and eventually   the local exponents
 $z$ corresponding to  $p=2$  become  larger.
 For $p=3$ and 4  the rate of increase  of $\ell_0$  remains smaller than for 
$p=2$ and 50 along the whole isotherm (for $|H|<0.008$).
According to the predictions summarized in Sec.~\ref{sec:2} for semi-infinite
systems $(L\to\infty)$,  upon
approaching the bulk coexistence line $H=0$ the thickness of the
wetting film 
is expected to diverge for $p\ge 3$ with the exponent $\beta_s^{co}=1/3$  and
for $p<3$ with the exponent $\beta_s^{co}=1/p$.
 As can be inferred  from  Fig.~\ref{fig:6}
the local exponents $z$  calculated for $p=50, 4, 3$, and 2 tend towards
 their predicted values for $H\to 0$ but cannot  reach them  due to  capillary
condensation.
For $p=1.5$ the local exponents  have  to behave non-monotonically in order to
reach the expected value $1/p=2/3$. The behavior of $\ell_0$
along the isotherm $T^*=1.9$ is very  similar (data not shown). 

Path (2)  corresponds to $T^*=2.0$ and 
 runs  in between the crossover lines given 
by $H_{cr}^{(1)}= -d_1|\tau|^{3\nu}$ 
and $H_{cr}^{(2)}=-d_2|\tau|^{\Delta}$ with $\nu=1$ and 
 $\Delta=15/8$ in $d=2$;
it  hits the pseudo-phase coexistence  line of capillary condensation
before it reaches the crossover boundary $H_{cr}^{(1)}=-d_1|\tau|^{3}$.
Along that path  the variation  of $\ell_0$  is very 
similar to the one observed along the isotherm $T^*=1.8$ with 
 the  local exponents
  approaching the expected values of  the complete wetting regime
(data not shown).
Again the asymptotic behavior is preempted by capillary condensation. 

\vspace{0.65cm}
\begin{figure}[h]
\centering
\includegraphics[width=7.5cm]{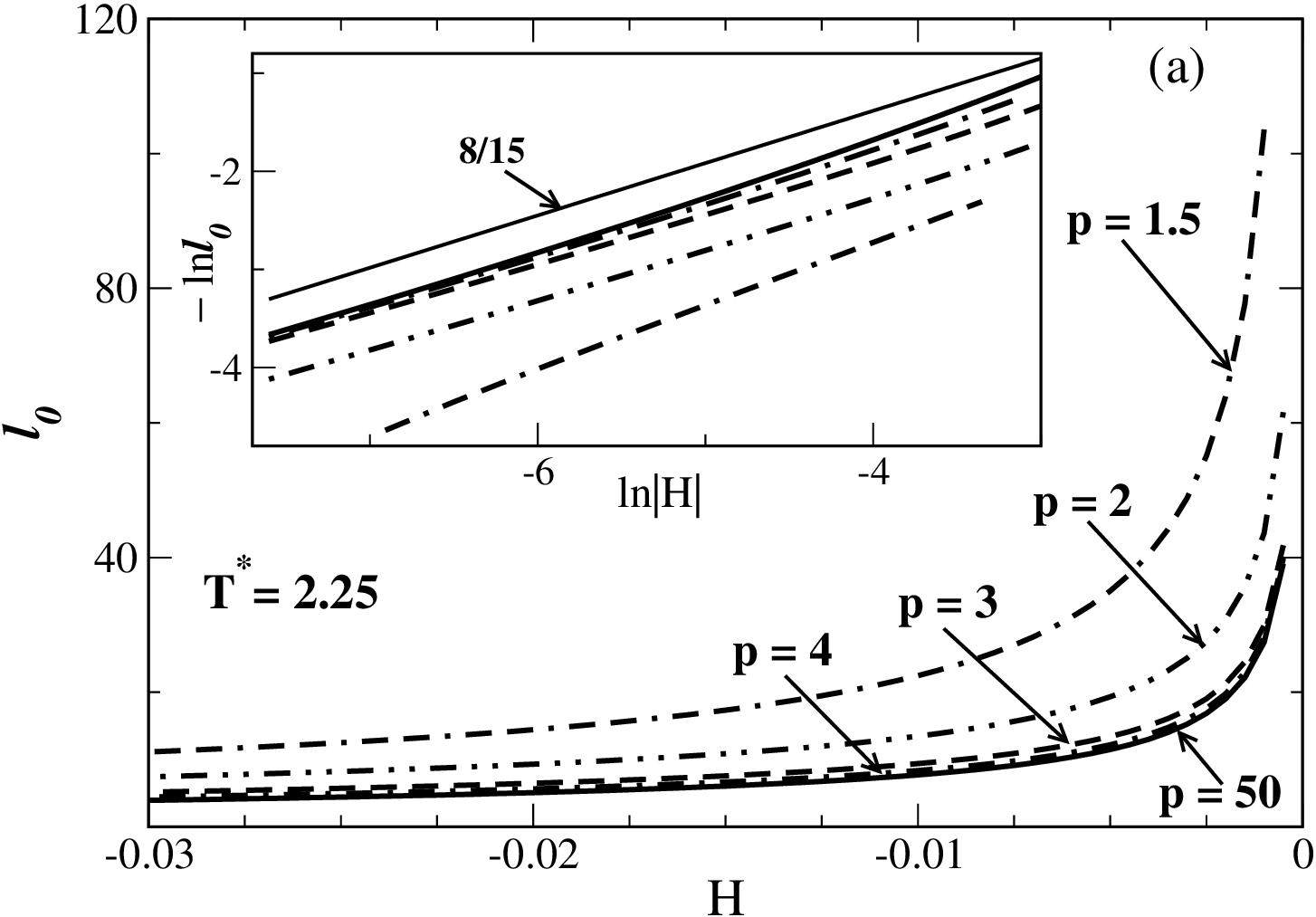}
\end{figure}
\vspace{0.08cm}

Results along the near-critical ($T^*=2.25$) and critical ($T^*=T^*_c$) isotherms
(paths (3) and (4) in Fig.~\ref{fig:1}) are displayed in Fig.~\ref{fig:7}
and  Fig.~\ref{fig:8}, respectively, for $L=500$. 
(For $L=300$ we observe a similar behavior; these  data are not shown.)
Both paths lie in the critical adsorption regime where one expects
$\ell_0(H\to 0) \sim \xi \sim |\Delta H|^{-\nu/\Delta}$ with $\nu/\Delta=8/15$ 
in $d=2$. This universal behavior is expected to
 hold if the long-ranged boundary field
decays sufficiently rapidly, i.e., if  $p>(d+2-\eta)/2$. 
In this case the long-ranged part of
the boundary field is irrelevant in the RG sense with respect to 
a pure contact surface field \cite{diehl}.  Here $\eta$ is the  critical exponent governing 
the algebraic decay of the two-point correlation function in the bulk and at $T_c$.
In the present case  of the $d=2$ Ising model $\eta(d=2)=1/4$ so that 
for $p> 15/8$ we expect to observe  the power law $\ell_0 \sim |\Delta H|^{-8/15}$.

For $p=1.5$ one has $p<(d+2-\eta)/2$. To our knowledge so far this case has not been
studied in the literature in the context of critical adsorption.
According to general scaling arguments \cite{diehl} the magnetization profile for the semi-infinite 
system under rescaling of distances by a factor $b$ behave as
\begin{align}
\label{eq:mlr}
 m_{\pm}(y,\tau,H,h_1,p) = &\mbox{} \nonumber \\
      & b^{-\beta/\nu}m_{\pm}(yb^{-1},\tau b^{1/\nu},Hb^{\Delta/\nu},h_1b^{\omega_s}), \nonumber \\
\end{align}
where $\omega_s\equiv (d+2-\eta)/2-p$. Because in the present case
$\omega_s>0$ the range of such wall-fluid
interactions is  relevant in the RG sense at and near $T_c$, i.e.,
the profiles are affected by the decay of the  boundary field even if 
 $y$ is sufficiently large; in the  critical region the profiles are no longer
equivalent to those generated by a pure contact surface field ($p=\infty$).
Choosing  $b\sim y$ one obtains the scaling form of the magnetization
\begin{equation}
\label{eq:mlr1}
m_{\pm}(y,\tau,H,h_1,p)=y^{-\beta/\nu}{\cal M}_{\pm}(y/\xi_{\tau},y/\xi_{H},y^{\omega_s}h_1),
\end{equation}
where $\xi_{\tau}\equiv
\xi_{\pm}(\tau,0)$ is given by Eq.~(\ref{eq:xitau})
whereas 
$ \xi_{H}\equiv \xi(0,\Delta\mu)$ is given by Eq.~(\ref{eq:ximu}).
For $\omega_s<0$  and  large enough $y$ the scaling function ${\cal M}_{\pm}$ 
can be expanded in terms of their third argument demonstrating  that in this case
the long range of the wall-fluid interaction  gives rise to  contributions
to the scaling functions of the magnetization profile which are subleading
 to the dominant universal $y^{-\beta/\nu}$ behavior valid  for
$y\lesssim \xi$ and $p=\infty$. 
Such an expansion cannot be performed for large $y$ if $\omega_s>0$
and in order to obtain a prediction in this case for the leading behavior 
of the magnetization profile  
and hence the thickness of the wetting layer $\ell_0$ and the adsorption,
 one has  to
calculate the full generalized scaling functions of the magnetization profiles. 
For $p<15/8$ such an analysis represents  a necessary future research goal 
which, however,  is beyond the scope
of the present study. Nevertheless, we find it instructive and stimulating
 to show our numerical results
for $p=1.5$ because they are strikingly different from those
obtained for  values of $p$ satisfying $p>(d+2-\eta)/2$.

As along the noncritical isotherms $T^* \le 2.25$ we find  that also   along the isotherms 
$T^*=2.25$ and $T=T^*_c$  the increase  of the wetting film thickness  upon  approaching 
$H=0$  is  much more stronger   for $p=1.5$ than for the  other values of  $p$. 
For $|H|\lesssim 0.01$ the local exponents attain  $z(p=1.5)\simeq 0.64$, which 
indicates an algebraic increase  of $\ell_0$ in this range of $H$ with 
$\ell_0\approx |H|^{-0.64}$. This effective exponent is close to the value
$1/p=2/3=0.66(6)$ predicted for the complete wetting behavior for $p=1.5$.
Closer to  $H=0$   the functional form of the increase  changes and  $z$ increases strongly. 

For $p=50, 4$, and 3  the   local exponents decrease upon approaching
the pseudo-phase coexistence line  $H_{ca}(T,L;p)$.  This behavior is  similar
to the one occurring along the isotherm  $T^*=1.8$;  however, apart from the case $p=1.5$ the  values
of $z$ in Fig.~\ref{fig:7}(b) are larger than the values in Fig.~\ref{fig:6}(b),  which indicates
the stronger increase of the wetting film thickness.
For $p=2$ the variation  of the local exponents differs from that for
the isotherm  $T^*=1.8$: here  $z$ is  a decreasing function of $H$,
almost identical to the one for $p=3$ (see Fig.~\ref{fig:7}(b)).

\vspace{0.65cm}
\begin{figure}[h]
\centering
\includegraphics[width=7.5cm]{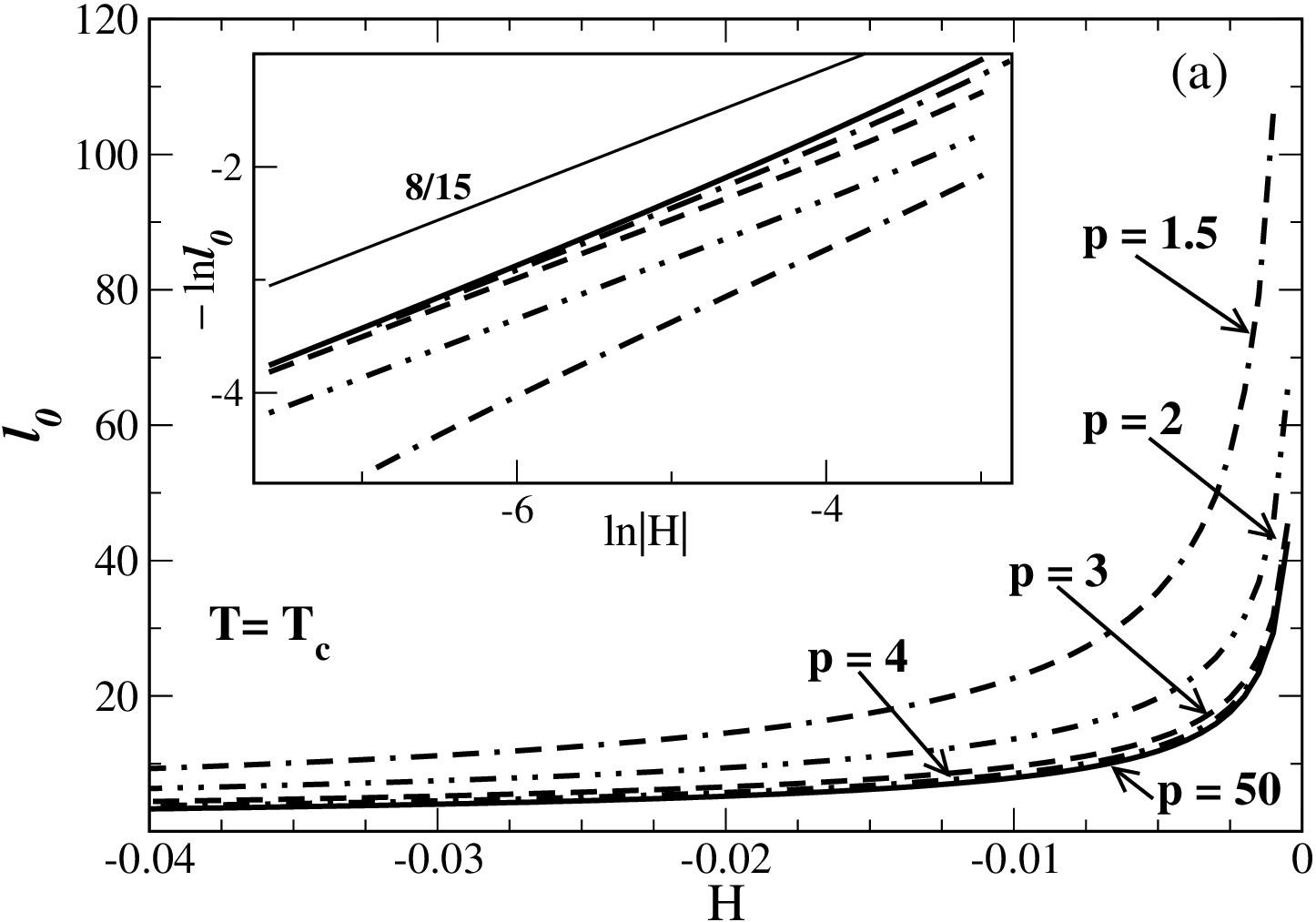}
\end{figure}
\vspace{0.10cm}

For all  decay exponents  $p\ge 2$ studied for $T^*=2.25$ the values of
the local exponents $z(H\to 0)$ seem to attain values between 0.48 and 0.52.
They would have to bend upwards for $H\to 0$ in order to reach the expected
value of  $8/15\simeq 0.533$. For the size $L=500$ capillary condensation
does not allow to reach the ultimate complete wetting regime below the
crossover line $H_{cr}^{(1)}$ (see Fig.~\ref{fig:1} for path (3)).

For  $p=3$ and 2, the variation of the  local exponents $z$ along the critical isotherm $T=T_c$
is very similar  to the one in Fig.~\ref{fig:7}(b). In both cases, $z$ first
decreases from the value $\approx  0.55$ above the expected one $8/15\simeq 0.53$ to the
value $\approx 0.52$ at $|H|\approx 0.001$ and only then, for even smaller
values of $H$, they seem to increase  towards the expected value $ 8/15$. In the limit $H\to 0$,
for  $p=50$ and 4 the local exponents seem to approach  the expected value
8/15 from above whereas for $p=3$ and 2 they will have to bend
upwards do so from below.

\vspace{0.55cm}
\begin{figure}[h]
\centering
\includegraphics[width=7.5cm]{rys10a.eps}
\end{figure}

\subsubsection{Adsorption}
\label{subsubsec:ads}

For the present microscopic model in the strip geometry the adsorption $\Gamma $ is defined as
\begin{equation}
\label{eq:adslatt}
\Gamma =\sum_{j=1}^{L/2}(m_{j}-m_b(T,H)),
\end{equation}
where $m_{j}\equiv \langle \sigma_{j}\rangle$ is the  magnetization
in row  $j$ and $m_b(T,H)$ is the corresponding bulk magnetization.
In the limit $L\to\infty$ this definition reduces to  the semi-infinite
quantity $\Gamma$ defined in  Eq.~(\ref{eq:1}). Along  the various
paths indicated in  Fig.~(\ref{fig:1}), at each point
we have first calculated the bulk magnetization $m_b(T,H)$
using an equivalent system but with a vanishing surface field in order
to minimize  the  influence of the boundary.

\vspace{0.65cm}
\begin{figure}[h]
\centering
\includegraphics[width=7.5cm]{rys11a.eps}
\end{figure}
\vspace{0.3cm}

\begin{figure}[h]
\centering
\includegraphics[width=7.5cm]{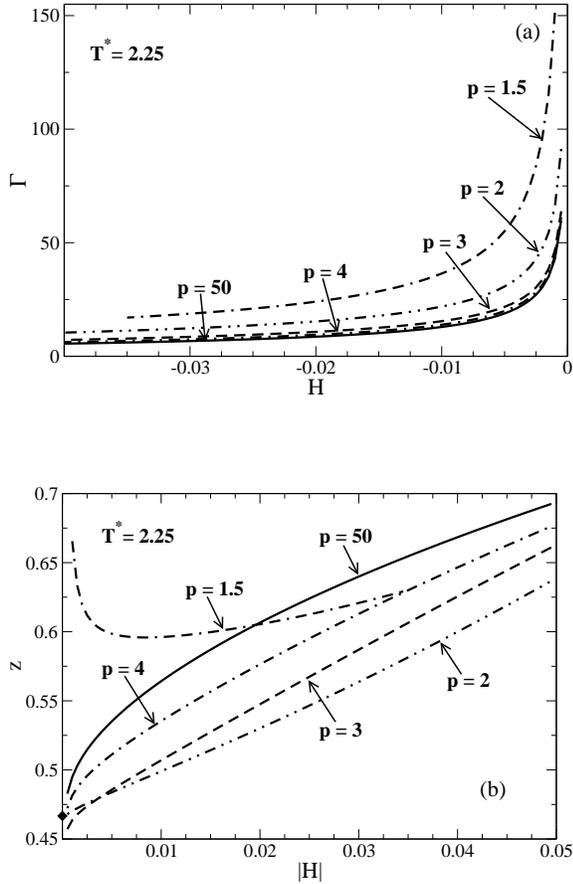}
\caption{The same as in Fig.~\ref{fig:9} along the isotherm $T^*=2.25$ (path (3) in Fig.~\ref{fig:1}).
In (b)  $(\nu-\beta)/\Delta=7/15$ is indicated $(\blacklozenge)$. The crossover to the ultimate
complete wetting behavior is masked by the occurrence of capillary condensation.
For $p=1.5$ the behaviors of $\Gamma(H)$ and $z(H\to 0)$ are distinctly different from those for $p\ge 2$.}
\label{fig:10}
\end{figure}

Moreover, for the latter calculations
as a representative spin the one in the middle of the strip was taken, as
it is the one least affected by finite-size effects. In order to obtain  the bulk
magnetization  $ m_b(T,H)$  we have extrapolated this  midpoint value
$m_{L/2}(T,H;h_1=0)$ to $L\to\infty$ from data calculated for $L=300, 400, 500, 600$, and $700$.

\begin{figure}[h]
\centering
\includegraphics[width=7.5cm]{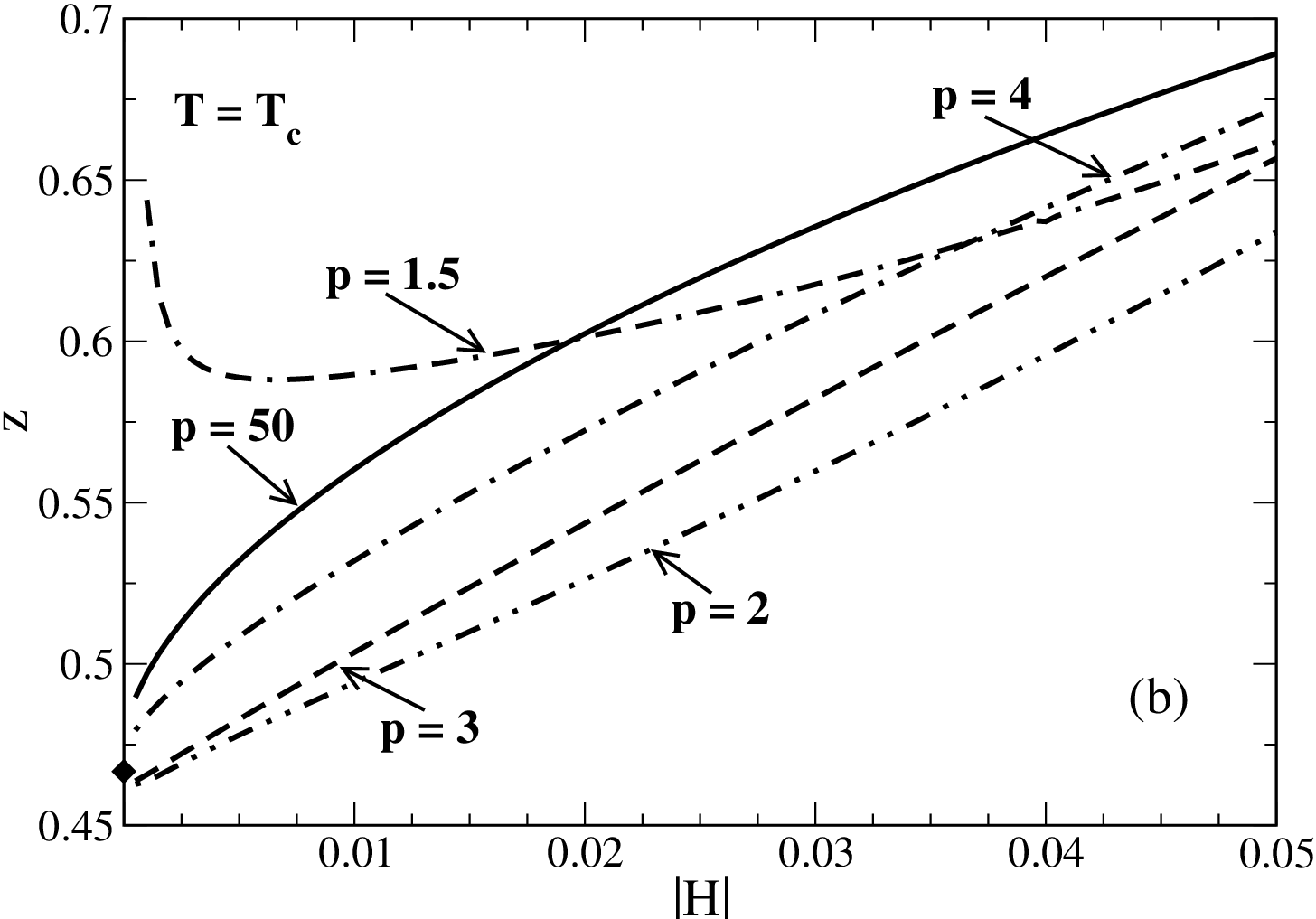}
\caption{ The same as in Fig.~\ref{fig:10} along the critical isotherm  $T=T_c$
(path (4) in Fig.~\ref{fig:1}).
In (b) $(\nu-\beta)/\Delta=7/15$ is indicated $(\blacklozenge)$. The crossover to the ultimate
complete wetting behavior is masked by the occurrence of capillary condensation.
For $p=1.5$ the behaviors of $\Gamma(H)$ and $z(H\to 0)$ are distinctly different from those for $p\ge 2$.}
\label{fig:11}
\end{figure}

The adsorption $\Gamma (H)$ and its local exponents calculated
for a  strip of  width $L=500$ with  $h_1=0.8$ and for various
ranges $p$ of the boundary fields are  shown in
Figs.~\ref{fig:9},~\ref{fig:10}, and~\ref{fig:11}  for the
paths (1), (3), and  (4), respectively, as indicated in Fig.~\ref{fig:1}.

\vspace{0.65cm}
\begin{figure}[h]
\centering
\includegraphics[width=7.5cm]{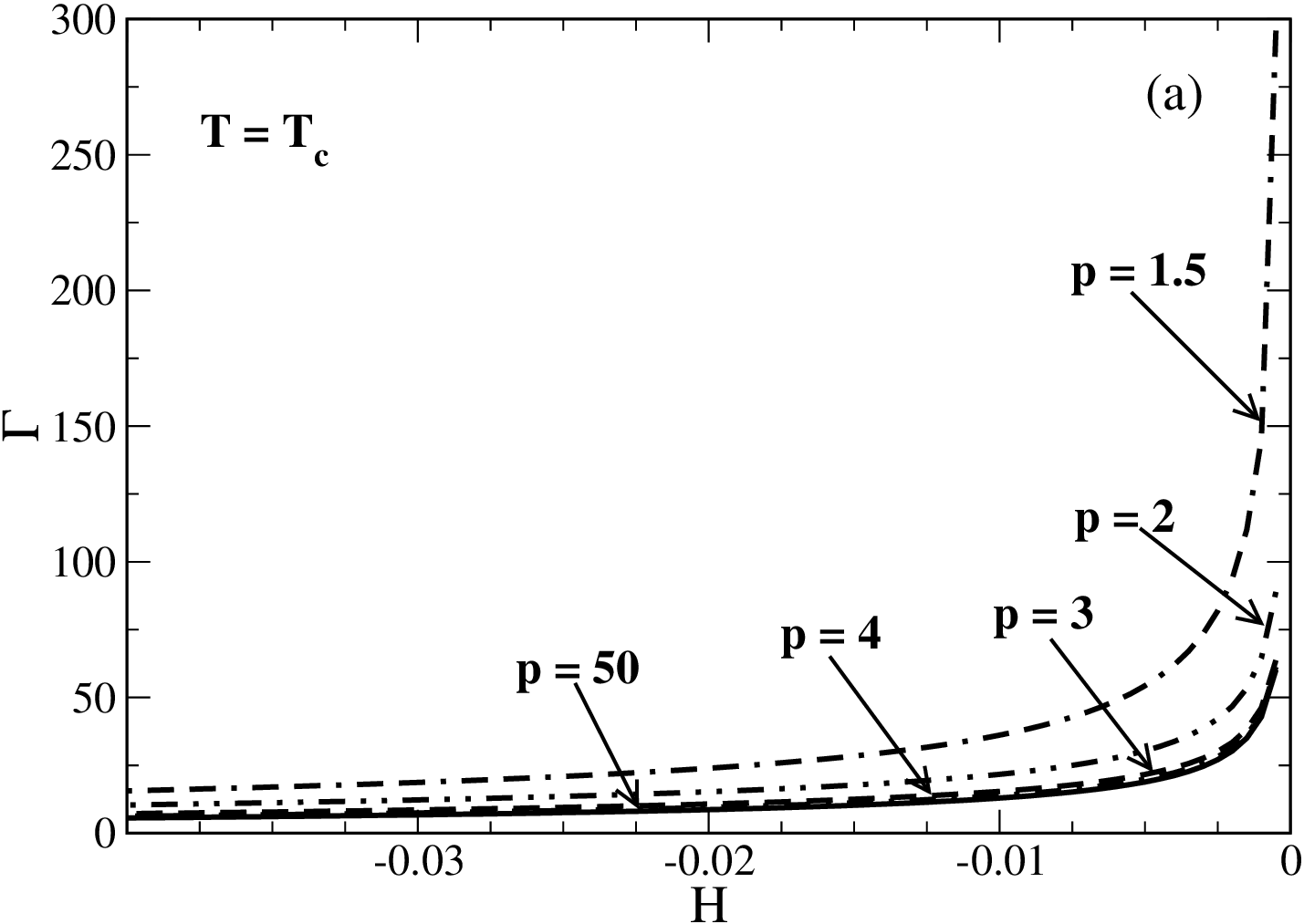}
\end{figure}
\vspace{0.15cm}

According to the predictions summarized in Sec.~\ref{sec:2},  upon
approaching the bulk coexistence line $H=0$  along
 path (1) $(T^*=1.8)$ the adsorption
is expected to diverge for $p\ge 3$ with the exponent $\beta_s^{co}=1/3$.
As can be seen in Fig.~\ref{fig:9}(b)
the local exponents calculated for $p=50, 4$, and 3
tend towards  this predicted
value for $H\to 0$ but   do  not reach it  due to the occurrence
of  capillary condensation.
For $p=2$ the local exponents as a function of the bulk field
 exhibit a negative curvature, consistent  with the behavior of $z$ for
the thickness $\ell_0$ of the wetting layer  (see previous subsection),
and  seemingly   tend to the exponent 1/2, which agrees
with the predicted non-universal behavior for $p<3$,
i.e., $\beta_s^{co}=1/p$. In order to reach the expected  exponent
 $\beta^{co}_s=2/3$ for $p=1.5$, the local exponents have to
vary in a  non-monotonic way  for $H\to 0$ (see Fig.~\ref{fig:9}(b)).

As expected from the analysis  of the thickness of the wetting layer,
the local exponents for the adsorption $\Gamma (H)$
along  path (2) ($T^*=2.0$) behave in a  similar way as along path (1) ($T^*=1.8$), i.e.,  
for $p=50, 4$, and 3  they tend to the complete wetting exponent $\beta_s^{co}=1/3$,
whereas for $p=2$ to the non-universal value $\beta_s^{co}=1/p=1/2$ (data not shown).
As before, the asymptotic behavior is not yet completely  reached 
when capillary condensation occurs.  We note that the theoretical considerations in Sec.~\ref{sec:2}
do not predict any specific power law behavior in terms of $H$ for the accessible $H$ values along this isotherm (2).

Paths (3) and (4) lie in the   critical adsorption regime and one expects (see Eq.~(\ref{eq:I_2}))
$\Gamma(H\to 0) \sim |H|^{-(\nu-\beta)/\Delta}$ as $H\to 0$  for $p\ge 15/8$.
One can see from  Figs.~\ref{fig:10}(b) and ~\ref{fig:11}(b) that for
 both $p=50$ and $p=4$  along both isotherms the local exponents seemingly  tend
 to the predicted value $(\nu-\beta)/\Delta=7/15$ from above.
In order to reach this value for $p=3$ and $2$,
 the local exponents  have to bend upwards for $H\to 0$;
 whether for path (3) they ultimately tend to 
the complete wetting value  $\beta_s^{co}=1/3$ and 1/2, respectively,
after crossing both crossover lines $H_{cr}^{(1)}$ and $H_{cr}^{(2)}$
is unclear because  for the finite values of $L$ considered here the
asymptotic behavior is undercut by capillary condensation. (Note that for $\beta_s^{co}=1/2$ one 
has $H_{cr}^{(1)}=d_1|\tau|^{2\nu}$.)
According to Fig.~\ref{fig:1} for $T^*=2.25$ this complete wetting regime is expected to be very narrow.

For $p=1.5$ $(<15/8)$ the behavior of the critical adsorption is expected to  be
non-universal. Indeed, as compared 
to the other values of $p$ we observe a distinctly  different   behavior of 
the local exponents along both paths (3) and (4) with    $z(H\to 0)$  
strongly increasing. (Note that for $p=1.5$ one has $H_{cr}^{(1)}=d_1|\tau|^{3\nu/2}$.)

For  $\tau\ne 0$ the capillary condensation line $H_{ca}(\tau,L)$ 
is expected \cite{parry:92} to approach the bulk coexistence line $H=0$ 
as  $1/L$ for $L\to \infty$ in accordance with the Kelvin 
equation  (Eq.~(\ref{eq:Kelvin})). Near $T_c$, i.e., for $|\tau|\ll 1$ 
scaling arguments lead  to the scaling behavior 
(see Eqs.~(\ref{eq:6}) and (\ref{eq:xitau})):
\begin{equation}
\label{eq:Kelvsc}
H_{ca}(\tau,L)= |\tau|^{\Delta}{\tilde g}(L/\xi,h_1|\tau|^{-\Delta_1}),
\end{equation}
 where $\Delta_1$ is the  surface gap exponent  $(\Delta_1(d=2)=1/2)$ due 
to the presence of surface fields ~\cite{diehl}; note that ${\tilde g}<0$.
(In Eq.~(\ref{eq:Kelvsc}) the scaling of $h_1$ holds 
if $p$ is sufficiently large,
 i.e., $p> 15/8$.) Since $\xi$ depends on $\Delta\mu$ (see Eq.~(\ref{eq:6})),
in Eq.~(\ref{eq:Kelvsc}) the first scaling variable of ${\tilde g}$
 depends also on $H_{ca}(\tau,L)$, rendering an implicit equation for $H_{ca}(\tau,L)$. To leading order 
$H_{ca}(\tau,L)$ is given by Eq.~(\ref{eq:Kelvsc}) with $\xi$
 from Eq.~(\ref{eq:xitau}).
In the following $h_1$ is considered to be large enough so that
${\tilde g}(x,u\to\infty)=g(x)<0$ becomes independent of the second scaling
variable.  The analytic property $g(x\to\infty)\sim 1/x$ renders
$H_{ca}(\tau,L\to \infty)\sim  |\tau|^{\Delta-\nu}/L $ for  $\tau\ne  0$
fixed and  $L\to \infty$, in accordance with the Kelvin equation.
On the other hand, for $L$ fixed and $\tau\to 0$ 
one has $g(x\to 0)\sim x^{-\Delta/\nu}$ (see below), 
which ensures that the limit
$\tau \to 0$ renders a nontrivial function of $L$:
 $H_{ca}(\tau\to 0,L)\sim L^{-\Delta/\nu}$ with 
$\Delta/\nu=15/8$  in $d=2$. For weak surface fields
 $H_{ca}(\tau,L)\sim h_1|\tau|^{\Delta-\Delta_1}{\bar g}(L/\xi)$ with 
${\bar g}(x\to 0)\sim x^{-(\Delta-\Delta_1)/\nu}$, ${\bar g}<0$,
 so that  $H_{ca}(\tau\to 0,L)\sim L^{-(\Delta-\Delta_1)/\nu}$ \cite{BLM}.

The capillary condensation line ends at a capillary critical point,
 at which the free energy of the confined system expressed in terms 
of the scaling variables
$(L/\xi,H|\tau|^{-\Delta})$ is singular at a  point $(x_0,y_0)$ 
implying Eq.~(\ref{eq:shifts}), which is consistent  with the scaling behavior
$H_{ca}(\tau,L\to\infty)\sim |\tau|^{\Delta-\nu}/L$. This scaling behavior
of the capillary condensation line is sketched  in Fig.~\ref{fig:12}.

\vspace{0.65cm}
\begin{figure}[h]
\centering
\includegraphics[width=8.0cm]{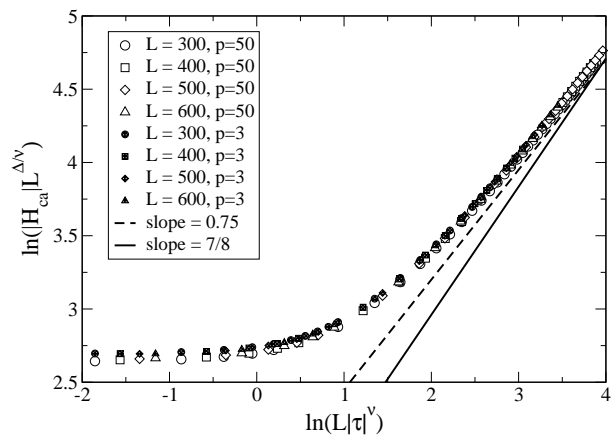}
\caption{The log-log plot of the scaling function ${\hat g}(x=L/\xi)$ (see Eq.~(\ref{eq:capsc})) of the
capillary condensation line $H_{ca}(\tau,L)$  calculated
for several large values of $L$ and  for a short-ranged ($p=50$)
and a long-ranged ($p=3$) boundary field. In the limit
$x\equiv L|\tau|^{\nu}\to 0$ the scaling function saturates at a constant
nonzero value whereas ${\hat g}(x\to \infty) \sim x^w$ with $w\simeq 0.75$.
The expected value $(\Delta-\nu)/\nu=7/8=0.875$
is not yet reached.}
\label{fig:13}
\end{figure}

With  $\xi \sim |\tau|^{-\nu}$ (see Eq.~(\ref{eq:6}))
the scaling form  for the capillary condensation line
can be written   equivalently as 
\begin{equation}
\label{eq:capsc}
H_{ca}(\tau,L)= L^{-\Delta/\nu}{\hat g}(L/\xi),
\end{equation}
 where
${\hat g}(x)=x^{\Delta/\nu}g(x)<0$. We test to which extent
 scaling holds  for the present model with $p=50$ and $p=3$
 by plotting $|H_{ca}|L^{\Delta/\nu}$ versus 
$L|\tau|^{\nu}$  for several strip widths $L$.
Although we neglect  the $H$ dependence 
of $\xi$ (see above and compare Eq.~(\ref{eq:6})) 
the data collapse is very good.
 From the log-log plot (see  Fig.~\ref{fig:13})
  one can see that ${\hat  g}(x\to 0)= const$
which  implies $g(x\to 0)\sim x^{-\Delta/\nu}$ (see above) and 
 $H_{ca}(0,L)\sim L^{-\Delta/\nu}$.
In the limit $x\gg 1$ our data fit very 
well to a  power law  ${\hat  g}(x \to \infty )\sim x^{w}$ 
with the effective exponent $w\simeq 0.75$.
$w$ is  smaller  than the value  $(\Delta-\nu)/\nu=7/8=0.875$
which would lead to the predicted behavior
 $H_{ca}(\tau,L\to \infty)\sim  |\tau|^{\Delta-\nu}/L $.
One might be inclined to put the blame
for the fact that $w$ has not yet reached this expected asymptotic value
on  corrections caused by the aforementioned $H$ dependence
of the bulk correlation length, according to which 
Eq.~(\ref{eq:capsc}) reads $H_{ca}(\tau,L) L^{\Delta/\nu}={\hat g}(L|\tau|^{\nu}(\Xi(H_{ca}|\tau|^{-\Delta})^{-1})$ 
(see Eq.~(\ref{eq:6})); this additional dependence actually spoils the scaling in terms of the scaling
variable $x=L|\tau|^{\nu}$. Since, however, scaling is observed, the latter
dependence must be weak and the too small value of the exponent $w$ must be due to other corrections.

\label{subsec:alon}

\vspace{0.65cm}
\begin{figure}[h]
\centering
\includegraphics[width=7.0cm]{rys14a.eps}
\end{figure}

\begin{figure}[h]
\centering
\includegraphics[width=7.0cm]{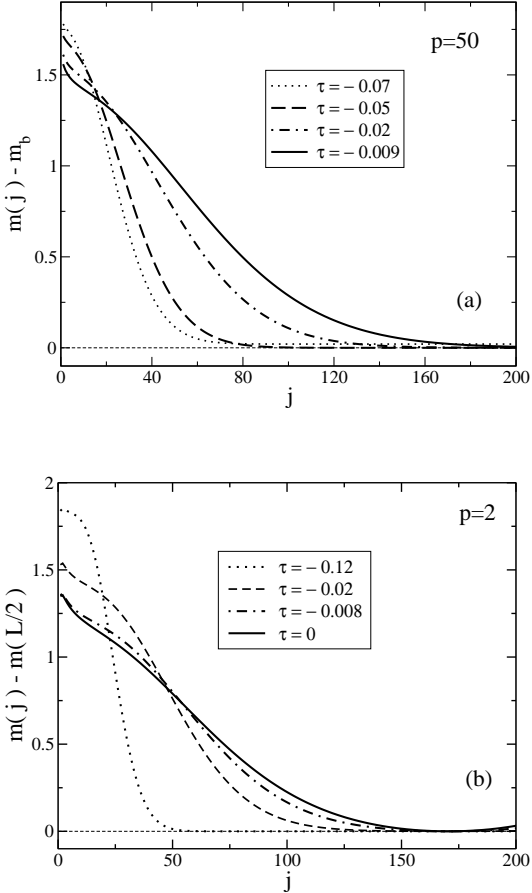}
\caption{Magnetization profiles near  one  wall, relative to their bulk values $m_b(T,H)$,
along  the thermodynamic path (0) (see Fig.~\ref{fig:1}) shifted by a constant $-10^{-6}$ away
from   the  pseudo-phase coexistence line $H_{ca}(T,L)$, calculated within
the DMRG method for  $d=2$ Ising strips of  width $L=700$
for short-ranged ((a)  $p=50$)  and  of width $L=340$ for
long-ranged ((b) $p=2$) boundary fields of strength $h_1=0.8$.
Due to the slow decay, the profiles for  $p=2$ are presented relative to their
values  $m(L/2)\ne m_b(T,H)$ at the midpoint of the film.}
\label{fig:14}
\end{figure}

Due to the scaling behavior $H_{ca}(\tau,L)\sim |\tau|^{\Delta-\nu}/L$, for
$L\gg 1$ but fixed, the pseudo-capillary condensation line appears to approach
the bulk critical point 
 (but without reaching it actually due to $L<\infty$) as
$|\tau|^{\Delta-\nu}=|\tau|^{7/8}$ 
which is  a  weaker   power law than $|H_{cr}^{(1)}|\sim |\tau|^{\nu/\beta_s^{co}}=|\tau|^3$ for 
the crossover line to  complete wetting. This explains the observation
that in $d=2$, upon approaching bulk coexistence sufficiently close to the critical point, the asymptotic 
complete wetting  regime is always preempted by capillary condensation. 
More generally, we expect that in  {\it confined} systems
the asymptotic complete wetting  divergence of the adsorption can be 
observed only if   $\Delta-\nu=2-\alpha-\beta-\nu > \nu/\beta_s^{co}$. 
This condition is fulfilled  neither in $d=2$ with $\Delta-\nu=0.875$ and $\nu/\beta_s^{co}=3$ nor in $d=3$
with  $\Delta-\nu=0.936$ and $\nu/\beta_s^{co}=3\nu=1.89$.
Thus we conclude that along an isotherm close to $T_c$ the ultimate crossover to the complete wetting 
behavior can only be observed in systems with macroscopically large transverse extensions.

In the present confined system the analogue of the thermodynamic path parallel to the bulk 
coexistence line  on the  gas side with a small undersaturation 
$\Delta \mu=const>0$ and with  $T\to T_c$ 
is the thermodynamic  path $H_{(0)}(T,L)=H_{ca}(T,L)+\delta H$
along the pseudo-phase coexistence (capillary condensation) line $H_{ca}(T,L)$
shifted   slightly by $\delta H$ to  the spin down ($\sigma=-1$) 
side  (see path (0) in Fig.~\ref{fig:1}).
In view of the large computational effort required for the determination of  the 
curve $H_{ca}(T,L)$ and of the value of the bulk magnetization $m_b(T,H)$
at each point along  this line we have calculated
 the adsorption along this path only for a  short-ranged boundary field
($p=50$), for the marginal case  $p=3$  of long-ranged surface fields,
and for $p=2$. We have considered three constant  shifts away
from the  pseudo-phase coexistence line: $\delta H=-10^{-5}, -10^{-6}$, and $-10^{-7}$. 

\subsubsection{Magnetization profiles}
\label{subsubsec:mp1}

In Fig.~\ref{fig:14} we have plotted a selection of 
magnetization profiles $m(z)$ along the thermodynamic 
path shifted by a constant $-10^{-6}$ away from the 
pseudo-coexistence line $H_{ca}(T,L)$ 
for $p=50$ (Fig.~\ref{fig:14}(a)) and for $p=2$ (Fig.~\ref{fig:14}(b)).
For  $p=3$ the profiles look qualitatively the same as  for $p=2$; 
they are both  characterized by a much broader  interfacial region  and larger
tails than the ones for $p=50$.
The shape of the  profiles  changes upon approaching  bulk coexistence 
$H=0$ in a way similar to the case  along the near-critical
isotherm $T^*=2.25$. One observes a broadening of the  interfacial region together
with  a rather small  region where the magnetization stays close to the value 
$m_b(T,H)>0$ characteristic for being slightly on  the upper side of the
bulk coexistence curve, i.e., $H = 0^{+}$. 

\subsection{Thermodynamic path along the pseudo -- phase coexistence  line}
\subsubsection{Thickness of the wetting layer.}
\label{subsubsec:wl1}

\begin{figure}[h]
\centering
\includegraphics[width=7.5cm]{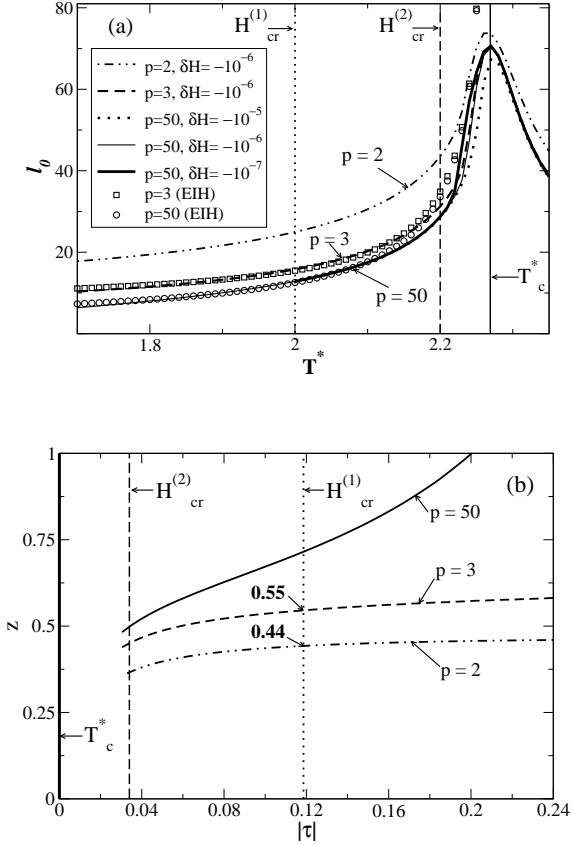}
\caption{(a) Wetting layer thickness $\ell_0$
(defined as in Figs.~\ref{fig:7} and \ref{fig:8}) as a function of
temperature  $T^*$
calculated for $L=340$, $h_1=0.8$, and
various values of the exponent $p$, characterizing the range of the
boundary fields, along the path (0) in Fig.~\ref{fig:1} with  constant  shift
$\delta H=  -10^{-6}$  away
from the pseudo-phase coexisting line $H_{ca}(T,L;p)$ (see main text).
For $p=50$, in addition  the temperature dependence of $\ell_0$ for
shifts  $\delta H= -10^{-5}$ and $-10^{-7}$ is  shown
(dotted  and thick solid line, respectively).
Symbols denote predictions from the effective interface Hamiltonian (EIH)
for $p=50$ (circle) and for $p=3$ (square) (see main text).
(b) The local exponents $z$ for $\ell_0$ calculated from
the data with  constant  shift
$\delta H=  -10^{-6}$ shown in (a) as a function of
$\tau=(T-T_c)/T_c$.  At the dotted and dashed vertical  line
the thermodynamic paths pass
through the crossover line  $H_{cr}^{(1)}$ and  $H_{cr}^{(2)}$, respectively,
shown in Fig.~\ref{fig:1} (and de facto independent of $\delta H$).
Due to the finite peak of $\ell_0$ at $T_c$, below $H_{cr}^{(2)}$, i.e., for
$|\tau|\lesssim 0.04$, the concept of the local exponents turns out to be
less useful so that the lines in (b) have not been continued into this region.}
\label{fig:15}
\end{figure}

For three undersaturations $\delta H$ relative to the pseudo-phase
coexistence line
Fig.~\ref{fig:15}(a) shows the thickness of the wetting layers
as a function of temperature calculated for a fixed strip width
$L=340$ with  strength $h_1=0.8$ and ranges $p=50, 3, 2$   of the
boundary fields. The increase  of the thickness of the wetting
layer along these  paths is very similar for all considered
values of the decay exponent $p$. 

\vspace{0.65cm}
\begin{figure}[h]
\centering
\includegraphics[width=7.0cm]{rys16a.eps}
\end{figure}

At low temperatures, for $p=2$  the wetting layer
is about three times thicker than for the short-ranged boundary field.
Except for the close  vicinity of the critical temperature,
i.e., for $-0.98\lesssim T^*-T^*_c \lesssim  1.01$,
where the finite-size effects are very pronounced,
results obtained along  paths with  different shifts  $\delta H$
but the same $p$ collapse onto a common curve. We note that for $p=50$ this common  curve
splits into three different lines just above
the pseudocritical temperature $T^*_{c,L}(L=340,h_1=0.8) \approx 2.2$, i.e.,
when the pseudo-coexistence line $H_{ca}(T,L)$ intersects the
crossover line  $H_{cr}^{(2)}=-d_2|\tau|^{\Delta}$.
For each $p$ the thickness of the
wetting film attains a certain $L$-dependent value at the bulk critical
temperature $T_c$, which also depends on the  value $\delta H$ of
the shift of the path (0).
In order to calculate $\ell_0(T=T_c)$ for large $L$ we have
performed  an extrapolation scheme for the curves $\ell_0(T)$ to $T=T_c$,
not taking into account data for temperatures higher than $T^*=2.24$ because those
exhibit  pronounced  finite-size effects beyond the leading $L$--behavior.
For $p=50$ we have found (data not shown) that  the thickness
of the wetting layer at $T=T_c$ is
proportional to  $L$, which is expected  because  $\ell_0$
cannot grow larger than $L/2$.  For $p\ge 2$ it should diverge for $L\to \infty$.
Above $T_c$ we observe a rapid decrease of $\ell_0$ as function of $T$.

For increasing $T\lesssim T_c$ in Fig.~\ref{fig:15}(a)
we observe a continuous  increase of the thickness of the wetting layer.
We recall that, according to   effective interface Hamiltonian studies,
 sufficiently far away from criticality, i.e., for $\ell_0 \gg \xi$, 
the wetting layer thickness should be independent of temperature along the path
of {\it constant}  undersaturation $\Delta \mu$, i.e.,  $ H=const$
with respect to bulk coexistence
(see the last paragraph of  Sec.~\ref{sec:2}).
The present
results  are calculated along  $H=H_{(0)}(L,T)=H_{ca}(T,L)+\delta H$ with $\delta H=const<0$
such that $|H|$ decreases for increasing
temperatures, which should result in  thicker films. This is in accordance 
with the data in Fig.~\ref{fig:15}(a).
In order to check whether  this  increase  is captured by
  effective interfacial models
we adopt  a simple effective interface potential $W(\ell)$ valid
for  quasi  short-ranged ($p=50$) boundary fields 
as well as for   long-ranged
boundary fields with a  decay exponent  $p=3$ or  2. Taking an  undersaturation
 $H$ into account, we make the ansatz \cite{dietrich} (in units of $k_BT_c/a^2$)
\begin{equation}
\label{eq:interfpot}
W(\ell)= A\ell^{-(p-1)}+ B\ell^{-\kappa}+|H|\Delta m\ell
\end{equation}
where an effective repulsive interaction $\sim B$, 
which  accounts for the gain in entropy
of the unbinding interface, has to be  added to the leading long-ranged
potential energy term $\sim A>0$ (see Subsec.~\ref{subsec:cw})~\cite{lipowsky}.
One has  $B=B_0 |k_BT/\Sigma(T)|>0$~\cite{lipowsky}, where $\Sigma(T)$ is 
the  surface stiffness of the interface, $B_0=const$ and $\Delta m=2|m_b^{(0)}(T)|$.
The exponent $\kappa$ describing thermal wandering is given by
$\kappa=2(d-1)/(3-d)$, which equals  2 in $d=2$ (see Subsec.~\ref{subsec:cw})
\cite{lipowsky}. 

In Eq.~(\ref{eq:interfpot}) the second (entropic) term
dominates in $d=2$ for $p>3$. Thus for  short-ranged forces $(p\gg 1$)
 the equilibrium wetting film thickness varies as function of $T$ along a thermodynamic
path $H(T)$ as 
\begin{equation}
\label{eq:lsrf}
\ell_0=(B_0k_BT/(\Sigma(T)|m_b^{(0)}(T)|))^{1/3}|H(T)|^{-1/3}, \quad p>3.
\end{equation}
For $p=3$ the entropic term  competes with the interaction term.
This leads to the equilibrium thickness of the wetting film
\begin{align}
\label{eq:llr}
\ell_0 = \;\;\;\;\;\;\;\;\;\;\:\;\;\;\;\;\;\;\;\;\;\;\;\;\;\;\;\;\;\;\;\;\;\;\;\;\;\;\;\;\;\;\;\;\;\;\;\;\;\;\;\;\;\;\;\;\;\;\;\;\;\;\;\;\;\;& \nonumber \\
((A+B_0k_BT/\Sigma(T))/|m_b^{(0)}(T)|)^{1/3}|H(T)|^{-1/3}, & \quad p=3. \nonumber \\
\end{align}
$A$ is proportional to $ \Delta m=2|m_b^{(0)}(T)|$ so that 
$A/|m_b^{(0)}(T)|$  only weakly depends on $T$. 
For $p=2$ the first term in Eq.~(\ref{eq:interfpot}) dominates 
and  the temperature dependence
of the film thickness follows that of $H(T)\simeq H_{ca}(T,L)$:
\begin{equation}
\label{eq:ellscl}
\ell_0=(A/(2|m_b^{(0)}(T)|))^{1/2}|H_{ca}(T,L)|^{-1/2}, \quad p=2.
\end{equation}

For the  $d=2$ Ising model the surface stiffness is known exactly
as  $\Sigma(T)/(k_BT)=\sinh 2(K-K^*)$
where $K=J/(k_BT)$ and $\tanh K^*=\exp(-2K)$.
Both $\Sigma(T)$ and $\Delta m(T)\equiv 2|m_b^{(0)}(T)| $ decrease
 with  increasing temperature.  Therefore, for all values of $p$,
$\ell_0$ is expected 
to increase along the pseudo-coexistence line $H_{ca}(T,L)$.

Sufficiently close to the bulk 
critical point the surface stiffness vanishes with the same power law
as the surface tension of the interface $\sigma$,
i.e.,  $\Sigma(\tau\to 0)\sim \sigma\sim |\tau|^{(d-1)\nu}=|\tau|$ 
 for the  $d=2$ Ising model.
Since $\Delta m \sim |\tau|^{\beta}$ it follows 
that in the critical region for short-ranged forces
\begin{align}
\label{eq:ellsc}
\ell_0\sim |\tau|^{-(\beta+(d-1)\nu)/3}|H_{ca}(\tau,L)|^{-1/3}=& \nonumber \\
|\tau|^{-3/8}|H_{ca}(\tau,L)|^{-1/3}, & \quad p=3. \nonumber \\
\end{align}
with $H_{ca}(\tau,L)$ given by Eq.~(\ref{eq:capsc}).

Close to $T_c$ and  in the temperature range for which 
 the pseudo-coexistence line $H_{ca}(\tau,L)$
lies in between the crossover lines  $H_{cr}^{(1)}=-d_1|\tau|^{3\nu}$
and   $H_{cr}^{(2)}=-d_2|\tau|^{\Delta}$ 
we have found (see  Fig.~\ref{fig:13}), for both  $p=50$ and $p=3$,
 that the absolute value of the  scaling function ${\hat g}(x=L/\xi)$
of the pseudo-coexistence line   (see Eq.~(\ref{eq:capsc}))
decreases and 
crosses over from  
 the power law behavior ${\hat g}(x\to \infty)\sim x^w$ with $w\simeq 0.75$
for  $|\tau|\ne 0$ to a constant 
value  ${\hat g}(x=0)$ at $\tau=0$. This implies that, for decreasing $|\tau|$, $\ell_0(\tau)$
crosses over from an increase $\sim |\tau|^{-5/8}$ to one $\sim |\tau|^{-3/8}$ (for $L$ fixed).
Thus the
effective interface  model predicts that, along the path (0),
 $\ell_0(\tau\to 0)$ is
an increasing  function of temperature. But in view of the aforementioned crossover 
one  cannot expect a
purely  algebraic behavior  in this region.

For $12 \lesssim L|\tau|^{\nu}\lesssim 50$,
 the variation of the pseudo-coexistence line $H_{ca}(T,L)$ 
 at fixed $L$ is approximately 
algebraic with an  effective exponent $w\simeq 0.75$.
Inserting $H=H_{ca}(T,L)\sim |\tau|^{0.75}$ into the expression for
$\ell_0$  (Eq.~(\ref{eq:ellsc})) gives 
$\ell_0 \sim |\tau|^{-0.625}$ for  both $p=50$ and $p=3$.
For $L|\tau|^{\nu}\ll 1$  one has  $H_{ca}\to const$, so that in this limit  Eq.~(\ref{eq:ellsc}) 
predicts  $\ell_0(\tau\to 0)\sim |\tau|^{-3/8}$.

For $p=2$  the pseudo-coexistence line $H_{ca}(T,L)$  
seems to follow  a power law with the same effective exponent 
as for $p=50$ and 3 for reduced  temperatures smaller than the
crossover line $H_{cr}(1)$ (see Fig.~\ref{fig:1}).  
Therefore on the basis of Eq.~(\ref{eq:ellscl}) for $12\lesssim L|\tau|^{\nu}\lesssim 50$
one  expects $\ell_0 \sim |\tau|^{-0.375}$.

For $p=50$ the behavior of the local exponents $z$ 
determined from the numerical  data obtained for $L=340$ and  shown  in 
 Fig.~\ref{fig:15}(a)  is consistent with the above
predictions  (see  Fig.~\ref{fig:15}(b)).  We find that 
 in between the crossover lines  $H_{cr}^{(1)}=-d_1|\tau|^{3\nu}$
and   $H_{cr}^{(2)}=-d_2|\tau|^{\Delta}$ the local exponents
$z$ decrease continuously upon decreasing $|\tau|$.
For $0.12 \lesssim L|\tau|^{\nu}\lesssim 50$, which with $L=340$ and $\nu=1$
corresponds to  $0.04 \lesssim |\tau|\lesssim 0.15$, the value of $z$ 
for $p=50$ changes from $\simeq 0.80$ to $\simeq 0.52$. No  
 saturation at the value 0.625, which would reflect  the
power law ${\hat g}(x)\sim x^{0.75}$ (see Eq.~(\ref{eq:capsc}) and the discussion above), is observed.
However, the above  temperature range lies outside the range of validity 
of the asymptotic
power law behaviors of $\Sigma(T)$ and $\Delta m(T)$ leading to  deviations
from the algebraic variation   $\ell_0\sim |\tau|^{-0.625}$ discussed above.

In order to test the validity of the effective interface Hamiltonian
approach we compare our full DMRG data with  the predictions for the increase of the
wetting film thickness along the line
of pseudo-phase coexistence  $H_{ca}(T,L)$ stemming from  Eq.~(\ref{eq:lsrf})
by  substituting our numerical data for
 $H(T)=H_{ca}(T,L)$ and  $\Delta m(T)=2|m_b(T)|$
and by using the analytic expression for $\Sigma/(k_BT)$
given above. Treating $B_0$ as a fitting parameter we find
very good agreement between these two approaches for $B_0=0.729$ (see Fig.~\ref{fig:15}(a))
in the wide temperature range below $T^*\simeq 2.215$.
We note, that for higher temperatures the system crosses
over to the critical adsorption regime (beyond the crossover line $H_{cr}^{(2)}$).

For $p=3$ and 2 we observe that at $H_{cr}^{(1)}$ the local exponents
$z$ saturate at the values $\simeq 0.55$ and  $\simeq 0.44$, respectively.
In the  case $p=3$ the exponent is slightly smaller than the
prediction 0.625 based on the scaling behavior obtained within the effective interface model,
whereas for $p=2$ it is slightly larger than the value 0.375 expected also within 
the effective interface model.

We repeat the procedure described above to check the agreement of the effective interface
Hamiltonian prediction in  Eq.~(\ref{eq:llr}) with our DMRG data for $p=3$. We adopt
$B_0=0.729$, i.e., the value determined from the fitting of Eq.~(\ref{eq:lsrf})   
to the DMRG results for $p=50$, and treat $A/|m_b(T)|\approx const\equiv C$ as a free parameter.
Very good agreement is obtained  for $C\approx 1.5$
and temperatures $T^*\lesssim 2.15$  (see Fig.~\ref{fig:15}(a)). 

However, the assumption $A/|m_b(T)|\approx const$ does not
lead to an agreement of the DMRG data for $p=2$ (not shown) with the curve
predicted by  Eq.~(\ref{eq:ellscl}) with $H(T)=H_{ca}(T,L)$. 
Thus we   conclude that in the case $p=2$  (not shown) the simple
 effective interface Hamiltonian given by Eq.~(\ref{eq:interfpot})
 fails to describe the full DMRG data.

\subsubsection{Adsorption}
\label{subsubsec:ads1}

\begin{figure}[h]
\centering
\includegraphics[width=7.5cm]{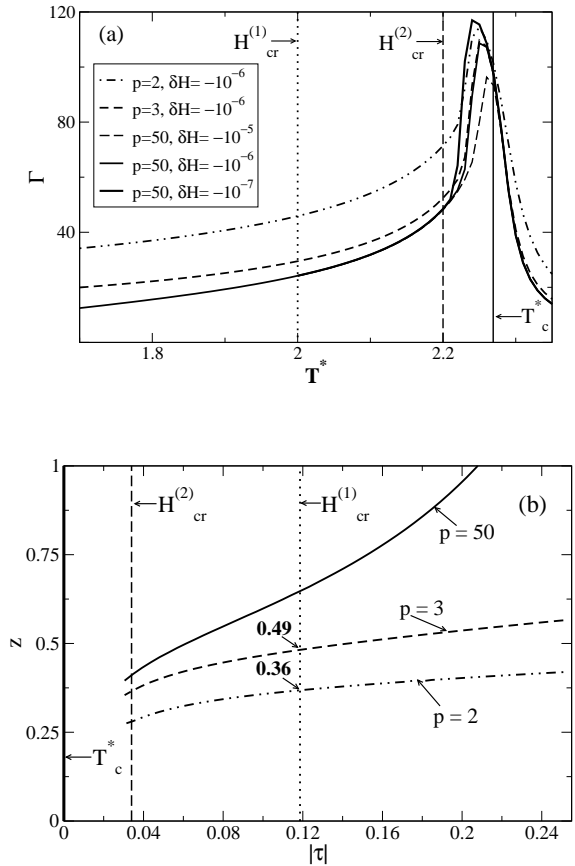}
\caption{ (a) Adsorption $\Gamma$ as a function of $T^*$ along
the path (0) in Fig.~\ref{fig:1} with a constant  shift
$\delta H= -10^{-6}$ away
from the pseudo-phase coexisting line $H_{ca}(T,L;p)$
(see main text), calculated within
the DMRG method for  $d=2$ Ising strips of  width $L=340$
for short-ranged ($p=50$) and long-ranged ($p=3,2$) boundary fields of
strength  $h_1=0.8$. For $p=50$ the temperature dependence  of $\Gamma$ for
shifts  $\delta H= -10^{-5}$ and $-10^{-7}$ is also shown (dotted and thick solid line, respectively).
(b) The local exponents $z$ for $\Gamma$ calculated from
the data  shown in (a) for the  constant  shift $\delta H=  -10^{-6}$.
 At the dotted and  dashed vertical line the thermodynamic path passes
through the crossover lines $H_{cr}^{(1)}$ and  $H_{cr}^{(2)}$, respectively,
shown in Fig.~\ref{fig:1} (and de facto independent of $\delta H$).
As in Fig.~\ref{fig:15}(b), in (b) the lines do not continue into the region
$|\tau|\lesssim 0.04$.}
\label{fig:16}
\end{figure}

Results for  the adsorption obtained for $L=340$,  $h_1=0.8$, and $p=50, 3, 2$ 
are shown as a function  of $T$ in Fig.~\ref{fig:16}(a).
\vspace{0.65cm}

\begin{figure}[h]
\centering
\includegraphics[width=7.0cm]{rys17a.eps}
\end{figure}
\vspace{0.1cm}

In accordance with  the expectations, the behavior of $\Gamma (T)$
upon increasing the temperature 
is qualitatively similar to the behavior of $\ell_0(T)$
in Fig.~\ref{fig:15}(a).
Results for  different  shifts $\delta H$ 
form a common curve above $T_c$  and for 
$T^*\lesssim 2.2$.
Close to $T_c$ the adsorption exhibits a maximum which signals
the vicinity of the end of the pseudo-phase coexistence line, i.e., 
the pseudo-critical point of a finite system. The position  of this  maximum
shifts towards $T_c$ for wider strips.
At $T_c$, $\Gamma$ attains a finite   value  $\Gamma (\tau=0,L)$.
As expected,  for long-ranged boundary fields the adsorption is slightly stronger
than for  short-ranged forces.

According to the discussion in the last paragraph of Sec.~\ref{sec:2},
for $p\ge 3$ and in between the two crossover lines
 $H_{cr}^{(1)}=-d_1|\tau|^{3\nu}$
and   $H_{cr}^{(2)}=-d_2|\tau|^{\Delta}$ (see Fig.~\ref{fig:1})
$\Gamma$ is expected to increase $\sim |\tau|^{-(\nu-\beta)}$,
which for $d=2$ corresponds to an exponent -0.875,
 provided the variation of $\xi$ is dominated by its dependence on
 $\tau$ (Eq.~(\ref{eq:xitau})).
This is the case  along  paths
of constant undersaturation $\Delta \mu \equiv H=const$.
Along the pseudo-coexistence line and  for $p=50$ the local exponents  
of $\Gamma$ 
 vary strongly  with  temperature, also below the crossover line
$H_{cr}^{(1)}$
(see  Fig.~\ref{fig:16}(b)). This is not compatible with the aforementioned
 algebraic variation of $\Gamma$ for $H=const$. 
In view of the discussion of the corresponding behavior 
of the wetting film thickness in the previous subsection, 
this difference can be traced back to the fact that the 
pseudo-coexistence line does not run in parallel to the bulk coexistence line.
 The decrease of the absolute value of  the bulk field 
$H_{ca}(T,L)$ upon  increasing temperature  matters and 
in the scaling analysis of $\Gamma$
 the whole  scaling form 
of $\xi$ (Eq.~(\ref{eq:6}))  has to  be taken into account.
According to Eqs.~(\ref{eq:5}) and (\ref{eq:6}) this is encoded in the product
of the scaling functions $\Xi_{\pm}$ and $K_{\pm}$ as function
of $H|\tau|^{-\Delta}$, which  alters the temperature dependence 
 $\sim |\tau|^{-(\nu-\beta)}$.
 Due to  $\Gamma \sim \ell_0\Delta m$  and the fact that $\Delta m$ decreases
upon decreasing the reduced temperature $|\tau|$, one expects that 
the increase of  
$\Gamma $ is weaker  than the increase of $\ell_0$. This
  is consistent with our numerical data for the local exponents $z$.
At each point along the  path (0) the value of $z$ obtained for
 $\Gamma$ is smaller
than the corresponding value of $z$ obtained for $\ell_0$  
(see  Figs.~\ref{fig:15}(b) and ~\ref{fig:16}(b)).

For $p=3$ and $p=2$ the variation of the local exponents of the adsorption 
in the  temperature range in between the crossover lines $H_{cr}^{(1)}$ and
$H_{cr}^{(2)}$ is less pronounced than that for $p=50$.
Near $H_{cr}^{(1)}$ we find saturation for the local exponents 
 $z$ at  values  which are 
smaller than  the saturation values  of $z$ 
 of the  corresponding thickness $\ell_0$
of the  wetting film, i.e., $z\simeq  0.49$
and  $z\simeq 0.36$ for $p=3$ and $p=2$, respectively
(see  Fig.~\ref{fig:16}(b)).
This is consistent with the relation  $\Gamma \sim \ell_0\Delta m$ and can be 
explained by the same arguments as used in the previous subsection.
We note, that  $\Delta m\sim |\tau|^{\beta}$, with 
$\beta=1/8=0.125$ for the $d=2$ Ising model,   leads  to  bigger
differences in the local exponents for $\ell_0$ and $\Gamma$ than 
the aforementioned  ones observed in the range of their saturation 
[$p=3$: $z(\ell_0)-z(\Gamma)=0.55-0.49=0.06<0.125$; $p=2$: $z(\ell_0)-z(\Gamma)=0.44-0.36=0.08<0.125$].
However, the above power laws are  satisfied only sufficiently close to $T_c$,
which is not reached along $H_{ca}(T,L)$ for the finite values of $L$ studied here.

\section{Summary and conclusions}
\label{sec:5}
We have studied two-dimensional Ising ferromagnets in  strip geometries
of width $L$ and  with long-ranged
boundary fields ($V^s_{j}=\frac{h_1}{j^p}, h_1>0$) (Eqs.~(\ref{eq:3_1})
and (\ref{eq:3_2})).
Based on scaling theory and the density-matrix renormalization-group 
method we  have obtained the following main results:

(1) In Sects.~\ref{sec:1} and \ref{sec:2} we have discussed the
 theoretical framework and the expectations
due to scaling theory  for the interplay between critical adsorption 
and complete wetting, including the crossover between them.
 For semi-infinite systems the corresponding scaling arguments lead to the schematic phase diagram shown in Fig.~\ref{fig:0}.

(2) The location of the
 pseudo-phase coexistence line $H_{ca}(T,L;p)$ of capillary condensation
 has been determined for various  ranges $p$ of 
the boundary fields (Fig.~\ref{fig:1}). For positive and 
parallel surface fields capillary condensation occurs at negative
 values $H_{ca}(T,L;p)$
of the  bulk field $H$. Increasing the range of the boundary fields
 shifts  the pseudo-phase coexistence line 
towards more negative values of the bulk field $H$
 (see Figs.~\ref{fig:1} and \ref{fig:2}). Relative to the pseudo-capillary
condensation line in the case that the surface fields act only on the two
boundary layers $j=1$ and $j=L$, i.e., for $p=\infty$, $H_{ca}(T,L;p)$ varies
exponentially as function of the range $p$ (Fig.~\ref{fig:2}).

(3)  A critical wetting transition has been identified for $p=2$, i.e., 
for a boundary field decaying slower than 
the marginal one  ($p=3$) (see Fig.~\ref{fig:3}). These findings
disprove previous claims by Kroll and Lipowsky \cite{kroll} that
in $d=2$  for $p<3$ there is no
critical wetting transition at finite values of $h_1$. 

(4) Within the accessible range of values for $T$ and $H$
the magnetization profiles both along  isotherms
(see Figs.~\ref{fig:4} and \ref{fig:5}) and along  the pseudo-capillary 
condensation line (see Fig.~\ref{fig:1})  are not slab-like.
 They exhibit a wide interfacial 
region and significant  tails. Primarily the profiles   approach
 their plateau values  in the middle of the strip exponentially 
as function of  the distance  from the wall. The width of the emerging 
interfacial region and the decay length of the  tails are proportional
to the bulk correlation length and thus  grow upon approaching
$T_c$ or the pseudo-capillary condensation  line. In the presence 
of  long-ranged boundary fields
the exponential decay of the profiles towards  their  bulk values $m_b(H,T)$
is followed by an algebraic  decay  $j^{-p}$ which finally is distorted
by the presence of the distant  wall. Features similar to those 
 of these order parameter  profiles
have been inferred from  neutron reflectrometry
 for the composition profile of a wetting film in a binary liquid mixture 
of n-hexane and perfluoro-n-hexane  \cite{bowers:07}.

(5) The variation of the thickness $\ell_0$  of the wetting layer 
along  various isotherms (Figs.~\ref{fig:6}-\ref{fig:8})
and along the pseudo-phase
coexistence line  (Fig.~\ref{fig:15}) has been analyzed for wide strips and different ranges
$p$ of the boundary field. 
Along both types of  path we have found a gradual increase of $\ell_0$
upon  approaching $T_c$.
The asymptotic divergence  of $\ell_0$  along the isotherms is preempted
by capillary condensation. As discussed on general grounds at the end of
Subsec.~\ref{subsubsec:ads}, both in $d=2$ and $d=3$ along the isotherm close to $T_c$ the ultimate crossover 
to the complete wetting behavior (see hatched region and path II in Fig.~\ref{fig:0}) can only be 
observed in systems with macroscopically large transverse extensions.
Along the isotherms and  within the  complete wetting regime the suitably defined  local exponents of
$\ell_0$ (Eq.~(\ref{zN})) tend  to approach the values predicted  by  the corresponding effective
interface Hamiltonian, i.e.,  $\beta_s^{co}=1/3$ for $p\ge 3$ and $\beta_s^{co}=1/p$ 
for $p<3$ (Figs.~\ref{fig:6} and \ref{fig:7}). These  results further support
our findings  that a critical  wetting transition exists also for the  values $p=2$ and $1.5$
of the decay exponent of the boundary field, i.e., smaller than the marginal case $p=3$.
Along the pseudo-phase coexistence line $H_{ca}(T,L)$ 
the variation of the thickness of the wetting film is determined
by the functional form of the temperature dependence of  $H_{ca}(T,L)$
and for $p>2$ it  agrees with the predictions of an  effective interface
Hamiltonian  (Eq.~(\ref{eq:lsrf}) and Fig.~\ref{fig:15}(a)).
For long-ranged boundary fields with  $p=3$ and $p=2$, the increase
of $\ell_0$ within a  certain range of temperature can be
described by a power law with an  effective exponent which is, however,
 not universal. It varies from $ca$ $ 0.55$ for $p=3$ to $ca$ $0.44$
for $p=2$ (Fig.~\ref{fig:15}(b)).

(6) For $p=50, 4, 3$, and $2$  the adsorption 
$\Gamma$   (Eqs.~(\ref{eq:1}) and (\ref{eq:adslatt})) 
has been calculated along various isotherms 
 (see Figs.~\ref{fig:9}-\ref{fig:11}). Along these thermodynamic 
paths, over  a  wide range of the bulk field $H$  the adsorption $\Gamma$ 
exhibits  a continuous  increase 
upon approaching the pseudo-phase coexistence line $H_{ca}(T,L;p)$.
Within the accessible range of values for $H$ this increase  cannot
be described by simple power laws.  Only very close to $H_{ca}(T,L;p)$
the local exponents of $\Gamma$  start to approach their
predicted values, i.e.,  $\beta_s^{co}(p\ge 3)=1/3$ and 
 $\beta_s^{co}(p=2)=1/2$
for  complete wetting (see Subsec.~\ref{subsec:cw} 
and Figs.~\ref{fig:9} and \ref{fig:10}))  and 
 $(\nu-\beta)/\Delta=7/15$ for  critical adsorption 
 (see Eq.~(\ref{eq:I_2}),
Subsec.~\ref{subsec:ca}, and Fig.~\ref{fig:11}).
However, these results are not entirely conclusive
 because the asymptotic regimes
are preempted by capillary condensation so that one is unable to
detect the crossover from critical adsorption 
to  complete wetting which is expected to occur sufficiently
close  to bulk coexistence for isotherms
with $T$ close to $T_c$. 
For the low temperature isotherms,
which lie entirely within the complete wetting regime, the cases  $p=1.5, 2$,
i.e.,  for $p$
smaller than the marginal value $p=3$, differ  distinctly
 from the  cases $p\ge 3$ (see Subsec.~\ref{subsec:cw} 
and Fig.~\ref{fig:9}), which is a clear indication of 
the non-universality of complete  wetting for these cases.
For  $p=1.5$ the long-ranged part of the boundary field is relevant in the RG sense and the behavior 
of $\Gamma$ is  non-universal along the critical isotherm (see Fig.~\ref{fig:11}(b)).

(7) The finite-size scaling predictions  for  the capillary condensation line
(see Eqs.~(\ref{eq:Kelvsc}) and~(\ref{eq:capsc}) as well as  Figs.~\ref{fig:12}
and \ref{fig:13}) are satisfied for both short-ranged  ($p=50$) and long-ranged
($p=3$) boundary fields. 

(8) For $p=50$, 3, and 2 the magnetization profiles (Fig.~\ref{fig:14})
and the  adsorption $\Gamma$ (Fig.~\ref{fig:16})
 have  been  calculated  along
the pseudo-capillary condensation line $H_{ca}(T,L)$. 
Over a  wide range of temperatures
below $T_c$,  $\Gamma$ increases gradually  (see Fig.~\ref{fig:16}).
Near criticality its increase cannot, however, be
described  by a simple  power law behavior. The reason is that 
along this thermodynamic path the temperature dependence of $\Gamma$
is determined also by the temperature variation of $H_{ca}(T,L)$, similarly
to the thickness $\ell_0$ of the wetting film which is approximately
related to $\Gamma$ as  $\Gamma \simeq \ell_0 \Delta m$ 
where $\Delta m(T)$  is the temperature dependent  difference 
between the  magnetization  of the
 spin up and the spin down bulk  phases. 
Our theoretical  results cannot be directly compared with  the experimental 
findings in Refs.~\cite{fenistein:02}
and  \cite{bowers:04}, in which  the thickness of the wetting layer
 \cite{fenistein:02} and the adsorption \cite{bowers:04}  were
measured along the paths corresponding to the path 
  (1) in Fig.~\ref{fig:0}, i.e., in these 
experiments the capillary condensation was absent.
The data obtained from  both experiments show that the 
behavior of $\ell_0$ and $\Gamma$ can be described by a power law 
but with an  effective exponent which  differs
from  both that for complete wetting and that for critical adsorption.
Similarly, in our theoretical analysis of two-dimensional Ising strips,
for long-ranged boundary fields with  $p=3$ and $p=2$ the increase
of $\Gamma$, for still sufficiently large reduced   temperatures $|\tau|$, can be
described by a power law with an  effective exponent, here  with a value of 
 $ca.$ $0.44$ for $p=3$  and of  $ca.$ $ 0.36$
for $p=2$  (Fig.~\ref{fig:16}). 

(9) Our results indicate that the asymptotic behavior for 
 $H\to 0$ occurs in a  much narrower regime
than one would expect and therefore requires much larger system sizes
to suppress capillary condensation sufficiently in order to reach the asymptotic regime.
 One indication for having entered the asymptotic 
regime could be that  a fully developed  wetting layer has been formed. For
the system sizes $L$ and thus the values of $H$ accessible in our calculations 
this does not yet fully occur.

\acknowledgments
A. D. thanks the Wroc\l aw Centre for Networking and
Computing (grant No. 82) and the Computing Center
of the  Institute of Low Temperature and Structure Research
PAS for access to their computing facilities.

\end{document}